\documentclass{article}
\usepackage{iftex}
\ifPDFTeX
  \usepackage[utf8]{inputenc}
\fi
\usepackage[T1]{fontenc}
\usepackage{arxiv}
\renewcommand{\shorttitle}{Econometric Anchoring and Cross-Market Attention}
\fancypagestyle{plain}{%
  \fancyhf{}%
  \fancyhead[C]{\shorttitle}%
  \fancyhead[R]{\scshape\footnotesize\headeright}%
  \fancyfoot[C]{\thepage}%
}
\usepackage{microtype}
\usepackage{amsmath,amssymb,bm}
\usepackage{graphicx,booktabs}
\usepackage[authoryear,round]{natbib}
\usepackage{placeins}
\usepackage[hidelinks]{hyperref}
\usepackage{url}
\graphicspath{{figures/}}
\makeatletter
\let\paperfloattable\table
\let\endpaperfloattable\endtable
\renewenvironment{table}[1][htbp]{%
  \def\papertablemode{#1}\def\paperheremode{H}%
  \ifx\papertablemode\paperheremode
    \par\addvspace{4pt}\noindent
    \begin{minipage}{\linewidth}\def\@captype{table}%
  \else
    \paperfloattable[#1]%
  \fi
}{%
  \ifx\papertablemode\paperheremode
    \end{minipage}\par\addvspace{4pt}%
  \else
    \endpaperfloattable
  \fi
}
\makeatother
\newcommand{\Needspace}[1]{%
  \par\begingroup
  \dimen0=\pagegoal\advance\dimen0 by -\pagetotal
  \ifdim\dimen0<#1\relax\newpage\fi
  \endgroup
}

\newcommand{\spplus}{\operatorname{softplus}}
\newcommand{\sigmoid}{\operatorname{sigmoid}}

\newcommand{\indicator}{\mathbf{1}}
\newcommand{\paperimage}[1]{\includegraphics[width=\textwidth]{figures/#1.pdf}}
\newsavebox{\papertablebox}
\newcommand{\densetable}{\fontsize{8}{9.2}\selectfont}
\newenvironment{papertabular}[1]{%
  \setlength{\tabcolsep}{3.5pt}%
  \begin{lrbox}{\papertablebox}%
  \begin{tabular}{@{}#1@{}}%
}{%
  \end{tabular}\end{lrbox}%
  \ifdim\wd\papertablebox>\linewidth
    \resizebox{\linewidth}{!}{\usebox{\papertablebox}}%
  \else
    \usebox{\papertablebox}%
  \fi
}
\title{Forecasting Global Equity Volatility: Econometric Anchoring and Cross-Market Attention}
\author{\textbf{Xinlin Zhao}\\
Independent Researcher\\China\\
\texttt{cillinzhao@gmail.com}
\And
\textbf{Haotian Qiao}\\
Department of Electrical Engineering\\and Computer Science\\
University of Michigan, Ann Arbor, USA\\
\texttt{qhaotian@umich.edu}}
\date{}

\begin{document}
\maketitle
\begin{abstract}
International volatility forecasting requires models to combine persistent local dynamics with information observed across different trading calendars. We propose PGA-Trans-HAR, an attention model that learns a bounded correction to a frozen heterogeneous autoregressive (HAR) forecast. A rolling predictive-connectedness graph guides spatial attention, while source masks accommodate market closures on the union calendar. We evaluate daily range-based volatility for eight equity indices from March 2013 to September 2026 at leads of 1, 5, and 22 union-calendar dates. Among seven primary forecasting models, PGA-Trans-HAR achieves the lowest equal-market average MSE at the 1- and 22-date leads and the lowest average QLIKE at all three leads. Its MSE reductions relative to HAR are 1.70\% and 1.13\% at the two endpoints. At the 1-date lead, the standard deviation of MSE across five fitted seeds is 0.00057, compared with 0.01260 for DCRNN-HAR and 0.03727 for GNN-HAR. The model is retained in 47 of 48 market-horizon MSE cells in the 90\% Model Confidence Sets across two evaluation panels. Under historical-tail calibration, joint VaR/ES scores improve on HAR at the 1- and 5-date leads. A common intraday exposure rule yields annual CER gains of 7.89--9.28 basis points relative to HAR across the specified cost scenarios. These results support bounded correction of a local econometric forecast as a design for accurate and stable cross-market volatility prediction.
\end{abstract}
\noindent\textbf{Keywords:} Volatility forecasting; HAR anchoring; Graph-guided attention;
Asynchronous markets; Forecast evaluation; Economic value.\\

\section{Introduction}
\subsection{Forecasting volatility across markets}
Volatility forecasts inform exposure decisions, risk limits, and the assessment of future financial uncertainty. International forecasting adds a second source of information: movements in one equity market can help predict volatility in another. The forecasting value of these links depends on how they are estimated and combined with local persistence. A useful model must learn from the network without allowing unstable relationships to dominate a well-established local benchmark.

The heterogeneous autoregressive model of \citet{corsi2009} provides a natural starting point. Its daily, weekly, and monthly components summarize persistence with few parameters. Multivariate extensions bring additional information, but also expand the number of estimated coefficients. For $N$ markets, a fully connected vector HAR specification contains $N(3N+1)$ coefficients. This expansion motivates structured information sharing, particularly when relationships vary across markets and forecast leads.

International trading calendars further complicate information sharing. A missing observation on a closed exchange is not an observation of zero volatility. Deleting all non-common dates removes usable market information, whereas unqualified zero padding changes the meaning of the data. The distinction between observed values and absent observations is central to asynchronous-data analysis \citep{burns1998}. It is equally important to separate the forecast origin from the target date. Cross-market graphs and neural inputs must be constructed from the history available at the origin, including when the forecast lead exceeds one date.

\subsection{A structured combination of econometric and neural forecasts}
Recent work develops complementary ways to combine financial information with flexible forecasting methods. \citet{kumar2025} combine variational mode decomposition with neural networks to forecast volatility. Their approach organizes the time series into components before nonlinear prediction. \citet{yu2026} extract cross-market spillover factors from realized covariance components and use them in machine learning forecasts of oil volatility. They also find value in forecast combinations. Together, these studies motivate a closer examination of how information processing and forecast aggregation shape the benefits of machine learning in finance.

Graph-based models provide a further route to cross-market information. \citet{zhang2025} study graph neural networks with spillover effects, while \citet{chi2026} combine dynamic graphs with all active trading days in a recurrent forecasting system. Attention learns input-dependent relationships \citep{vaswani2017}. Building on these approaches, we ask whether a bounded neural correction to HAR can improve forecasting accuracy and fitting stability across asynchronous markets. The econometric forecast remains fixed during neural training, and the network learns an adjustment with an explicit amplitude bound. The rolling graph is estimated from information available at each forecast origin, while source masks govern which market observations enter attention. The resulting system links the availability of international information to the size of its adjustment to a local forecast.

We address this question through a structured forecasting system built around HAR. PGA-Trans-HAR has two paths. The econometric path produces a horizon-specific OLS HAR forecast. The neural path combines temporal attention with spatial attention guided by a rolling generalized forecast error variance decomposition (GFEVD) graph. A learned gate for each destination market controls the mixture of graph information and data-driven attention. The neural output enters as a bounded correction to the frozen HAR forecast. A positive output map preserves the volatility scale.

This design assigns distinct tasks to distinct components. The HAR anchor supplies persistent local dynamics. The rolling graph summarizes predictive connectedness over a longer trailing window. Attention uses the current input representation to adapt information aggregation. Market-specific gates retain a compact allocation structure, with one fitted parameter per market. Finally, a node-balanced objective gives each observed market a direct role in training and emphasizes large relative errors against its own HAR reference.

\subsection{Contributions and empirical findings}
The primary contribution is an econometric anchoring mechanism for cross-market neural volatility forecasting. It conserves a fitted HAR forecast and learns a bounded adjustment in the inverse-softplus domain. This gives the neural path a well-defined task: refine the local forecast using international information. A second contribution is the integration of an origin-aligned rolling graph with source-masked attention. The graph and masks make cross-market inputs usable on the union calendar, including when exchanges close on different dates. A third contribution is the evaluation of this anchored system across forecast leads, markets, and fitted seeds. Market-specific gates provide one compact allocation rule; fixed and shared-gate variants assess the role of allocation flexibility within the same framework.

The sample contains eight international equity indices from 2013 to 2026. We use a common data source and measure daily volatility through the high--low range. Models are evaluated at leads of 1, 5, and 22 union-calendar dates. The assessment includes six external benchmarks, six controlled variants, two target-date panels, and five independently fitted neural seeds. Reliability is assessed through forecast losses, variation across fitted seeds, and performance across markets and leads. Pairwise inference and Model Confidence Sets complement these comparisons. Intraday VaR/ES scores and a cost-sensitive exposure experiment provide supporting evidence on forecast use.

The chief finding combines incremental accuracy with stable fitting. Among the primary models, PGA-Trans-HAR has the lowest average MSE at the 1- and 22-date leads and the lowest average QLIKE at all three leads. At the 5-date lead, its MSE remains close to HAR, alongside improvements in average MAE and QLIKE. Short-lead improvements against HAR are statistically significant in several markets. The model is retained in 47 of 48 MSE confidence-set cells across the two panels. Its MSE also varies less across fitted seeds than those of GNN-HAR and DCRNN-HAR at each evaluated lead. The controlled variants show that simple allocation rules work well within the anchored architecture. These findings make the combination of econometric anchoring and structured information sharing the main empirical result. The subsequent risk and exposure exercises examine how that forecasting performance carries through to particular decision rules.

Section~2 presents the forecasting foundations and benchmarks. Section~3 describes the model. Section~4 explains the data and evaluation protocol. Section~5 reports forecast accuracy and inference. Sections~6 and~7 examine tail risk and economic value. Section~8 concludes.

\section{Theoretical foundations and benchmark models}
\subsection{HAR and multivariate extensions}
Let $y_{t,n}>0$ denote observed daily volatility for market $n$. A direct HAR forecast at lead $h$ takes the form
\begin{equation}
 y_{o+h,n}=\beta_{0,n}^{(h)}+\beta_{d,n}^{(h)}d_{o,n}
 +\beta_{w,n}^{(h)}w_{o,n}+\beta_{m,n}^{(h)}m_{o,n}
 +\varepsilon_{o+h,n},
 \label{eq:har}
\end{equation}
where $o$ is the forecast origin. The three regressors summarize the most recent valid value and masked averages over 5 and 22 union-calendar positions. Thus, a five-position average can contain fewer than five observed market days. This convention keeps the input calendar consistent across models. Each horizon and market has its own OLS coefficients.

VHAR uses the three HAR regressors from all markets in every destination equation. HAR-KS uses the destination's three regressors and the latest valid daily value from each other market. In the eight-market panel, the resulting coefficient counts, including intercepts, are 200 for VHAR, 88 for HAR-KS, and 32 for separate HAR equations. These models distinguish the value of unrestricted cross-market regressors from the value of structured network aggregation.

\subsection{Graph-based forecasting benchmarks}
GNN-HAR follows the graph-based forecasting motivation of \citet{zhang2025}. In our implementation, a graphical lasso estimates a pretest precision matrix \citep{friedman2008}. Absolute partial correlations, with self-links, form a row-normalized graph. A two-layer graph convolution uses three summaries: the latest valid observation in the 22-position input window, a masked mean over positions \(o-4,\ldots,o-1\), and a masked mean over positions \(o-21,\ldots,o-5\). These inputs provide a daily reference and two preceding calendar summaries. The graph output is combined with a trainable HAR component and mapped through softplus. This HAR component uses the overlapping 5- and 22-position averages defined above. The graph is re-estimated on the refit sample and held fixed during testing.

DCRNN-HAR follows the dynamic graph and asynchronous-calendar framework of \citet{chi2026}. Its encoder and autoregressive decoder use diffusion-convolutional gated recurrent units. The graph is recomputed at each origin from the 22-position input history using VAR(1) and GFEVD. The decoder produces the intermediate forecasts up to the requested lead and feeds back predicted values rather than realized future volatility. A trainable HAR component contributes to the output using the same daily reference and preceding-calendar summaries as the GNN graph inputs. Training uses observed intermediate targets, while validation and reported evaluation concern the requested endpoint. The decoder receives a future activity mask constructed from observed data availability and supplied under a known-calendar assumption. At each decoder step, the previous prediction and the activity mask determine the next input and graph aggregation.

These benchmarks represent network information through a static undirected graph for GNN-HAR and an origin-specific directed graph for DCRNN-HAR. The primary comparison evaluates their complete forecasting systems. The controlled variants isolate components within the proposed architecture.

\subsection{Attention without the econometric paths}
Pure-ST-Transformer is the data-driven attention benchmark. It uses masked temporal and spatial attention, market and position embeddings, and a positive output head. It omits both the rolling GFEVD prior and the frozen HAR anchor. Its width, number of blocks, and temporal heads match the main attention path. Comparing this model with the full system assesses the value of the combined econometric restrictions. The controlled variants in Section~4 isolate changes within the anchored architecture.

\section{PGA-Trans-HAR: the forecasting architecture}
\begin{figure}[tbp]
\centering\paperimage{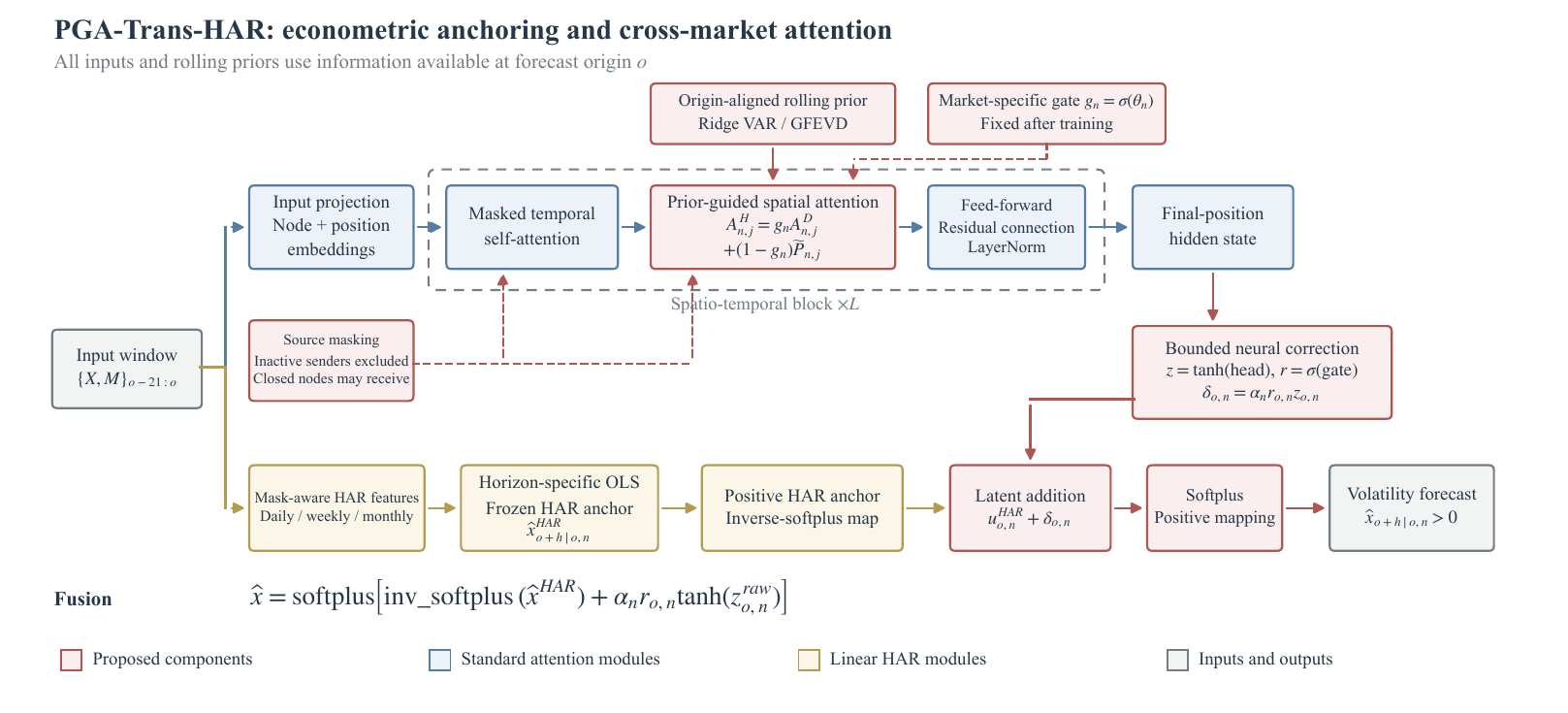}
\caption{Model architecture and econometric anchoring. The neural path aggregates a 22-position input window using masked temporal attention and graph-guided spatial attention. A fixed gate for each destination market weights data-driven attention and the rolling predictive-connectedness prior. The parallel econometric path produces a frozen, horizon-specific HAR forecast. A bounded correction enters its inverse-softplus representation before the forecast is mapped back to positive volatility.}
\label{fig:architecture}
\end{figure}

\subsection{Information set and the rolling predictive graph}
Index union-calendar dates by $t=0,\ldots,T-1$ and define $o=t-h$. The daily information set is
\begin{equation}
 \mathcal F_o=\sigma\{y_{s,n},M_{s,n}:s\leq o,\ n=1,\ldots,N\},
 \qquad M_{s,n}=\indicator\{y_{s,n}\text{ is valid and observed}\}.
 \label{eq:information}
\end{equation}
The neural window is $[o-S+1,o]$, with $S=22$. Missing entries are numerically represented by zero and accompanied by $M$. The zero is a storage value, not a volatility observation. The target mask excludes unobserved outcomes from the loss.

The predictive graph is constructed from a trailing window ending no later than $o$. To make the update rule reproducible, let $u=o+1$ be the exclusive endpoint and set
\begin{equation}
 e(o)=S+K\left\lfloor\frac{u-S}{K}\right\rfloor,
 \qquad a(o)=\max\{0,e(o)-W\},\quad W=252,\quad K=20.
 \label{eq:graphcalendar}
\end{equation}
The graph uses observations in $[a(o),e(o))$, so that its last possible observation is $e(o)-1\leq o$. Each graph is reused for 20 consecutive exclusive origin endpoints. The early estimation windows can be shorter than 252 positions. The refresh schedule is indexed by the origin rather than the target, making the rule identical across forecast leads.

For graph estimation, missing values are forward-filled within this trailing window. Remaining leading gaps use the within-window market median, with a zero fallback if the entire series is absent. This completion is confined to graph estimation; the neural path retains its original source masks. Each completed series is centered and scaled within the same window. A ridge VAR(2) is then fitted by
\begin{equation}
 \widehat B=(Z'Z+\rho D)^{-1}Z'Y,
 \qquad \rho=5,\quad D=\operatorname{diag}(0,1,\ldots,1),
 \label{eq:ridge}
\end{equation}
where $Z$ contains the intercept and lagged standardized values. A pseudoinverse is used if the linear solve is singular. Shrinkage controls unstable coefficient estimates \citep{hoerl1970}. The innovation covariance uses the residual cross-product divided by the number of regression observations, with a $10^{-6}$ diagonal stabilizer.

Following generalized impulse-response analysis \citep{pesaran1998} and predictive connectedness measurement \citep{diebold2012}, let $\Psi_k$ be the moving-average coefficients and $\Sigma$ the estimated innovation covariance. The generalized variance share is
\begin{equation}
 \Theta_{ij}(H_G)=
 \frac{\Sigma_{jj}^{-1}\sum_{k=0}^{H_G-1}
 (\bm e_i'\Psi_k\Sigma\bm e_j)^2}
 {\sum_{k=0}^{H_G-1}\bm e_i'\Psi_k\Sigma\Psi_k'\bm e_i},
 \qquad
 P_{ij}(o)=\frac{\Theta_{ij}(H_G)}{\sum_j\Theta_{ij}(H_G)},\quad H_G=22.
 \label{eq:gfevd}
\end{equation}
Rows are destinations and columns are sources. The GFEVD horizon $H_G$ is distinct from the forecasting lead $h$. Own-market shares are retained. The resulting graph provides a predictive allocation of variance shares, without requiring an ordering of the markets.

\begin{figure}[tbp]
\centering\paperimage{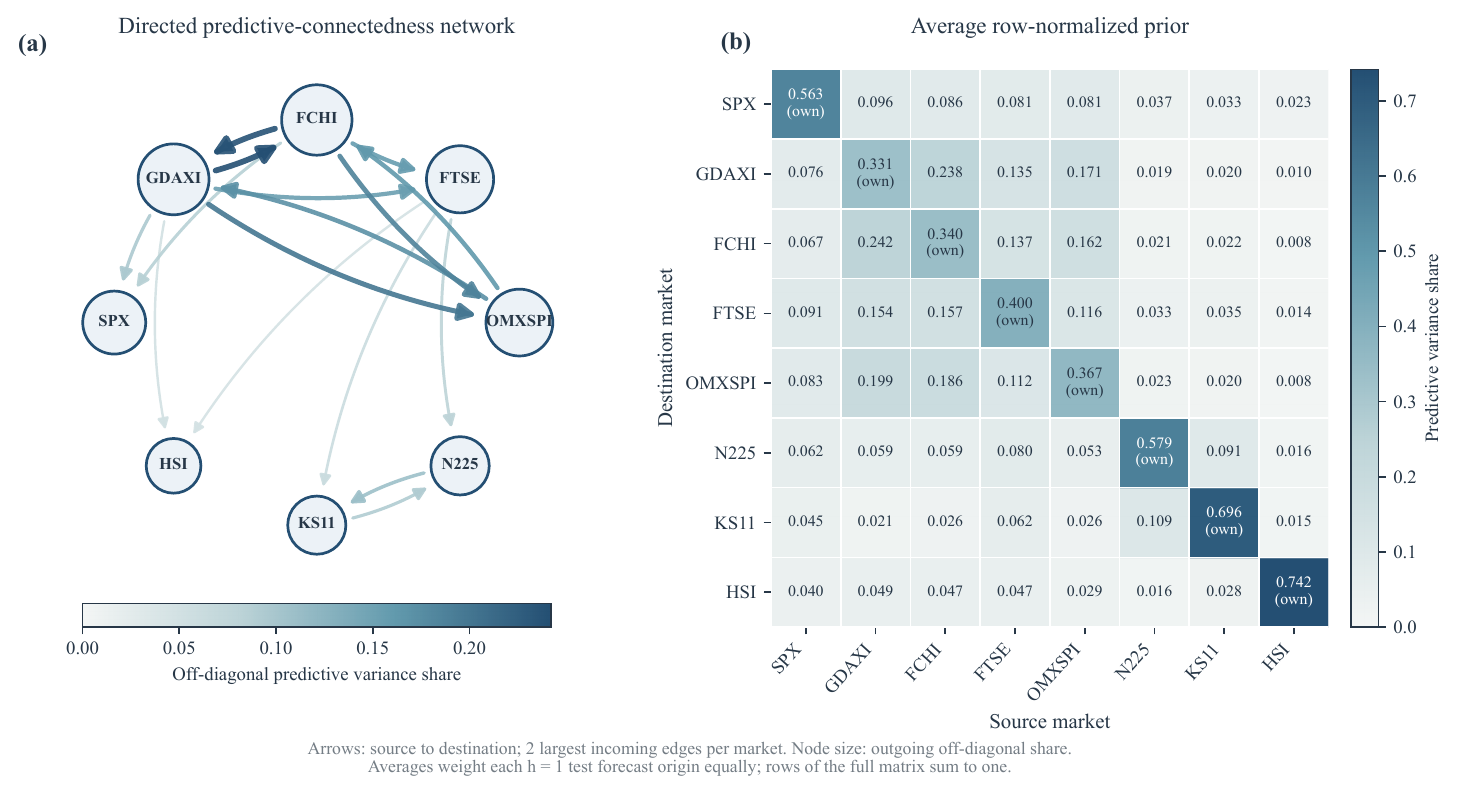}
\caption{Rolling predictive-connectedness prior. Panel (a) shows the two largest off-diagonal incoming links for each destination. Arrows point from source to destination; edge width and colour represent average predictive variance shares. Node size reflects outgoing off-diagonal shares. Panel (b) displays the full row-normalized matrix, including own-market shares. Both panels average the prior over the $h=1$ test origins with equal origin weights. Each component graph is estimated from its own admissible trailing history.}
\label{fig:prior}
\end{figure}
Figure~\ref{fig:prior} summarizes the network supplied to the attention path. The row-normalized representation separates self-dependence from cross-market inputs and preserves directional asymmetry. The figure averages the sequence of origin-specific priors. Each forecast uses the matrix assigned to its own origin.

\subsection{Mask-aware temporal attention}
Each input pair $(y_{s,n}M_{s,n},M_{s,n})$ is projected to a 32-dimensional hidden state and augmented with learned market and position embeddings. Temporal attention operates separately for each market. For query position $s$, head $k$, and key position $r$, the attention weight is
\begin{equation}
 A^{T,k}_{s,r,n}=
 \frac{M_{r,n}\exp(q_{s,n}^{k\prime}k_{r,n}^k/\sqrt{d_k})}
 {\sum_{v=o-S+1}^{o}M_{v,n}\exp(q_{s,n}^{k\prime}k_{v,n}^k/\sqrt{d_k})}.
 \label{eq:temporal}
\end{equation}
A stable masked softmax implements this expression; an empty key set returns zero aggregation. There are four temporal heads. No additional triangular mask is needed because every input position is already at or before the forecast origin. A closed-date query can still aggregate valid earlier keys. This maintains a representation at the calendar endpoint even when that market was most recently closed.

\subsection{Market-specific prior-guided spatial attention}
At each input position, data-driven spatial attention compares market states. If $i$ is the destination and $j$ the source,
\begin{align}
 A^D_{s,ij}&=
 \frac{M_{s,j}\exp(q_{s,i}'k_{s,j}/\sqrt d)}
 {\sum_vM_{s,v}\exp(q_{s,i}'k_{s,v}/\sqrt d)},\label{eq:dataattention}\\
 \widetilde P_{s,ij}(o)&=
 \frac{M_{s,j}P_{ij}(o)}{\sum_vM_{s,v}P_{iv}(o)},\label{eq:maskedprior}\\
 A^H_{s,ij}&=g_i A^D_{s,ij}+(1-g_i)\widetilde P_{s,ij}(o),
 \qquad g_i=\sigmoid(\vartheta_i).\label{eq:hybrid}
\end{align}
If a masked prior row has zero mass, it falls back to the data-attention row. When all sources are absent, the aggregation is zero. The source mask is asymmetric: an absent market cannot transmit through the attention weights, but a destination state is not erased. The convex combination in Equation~\eqref{eq:hybrid} is row-normalized before attention dropout.

Exactly one logit $\vartheta_i$ is fitted for each destination. It is initialized at $-0.5$, corresponding to $g_i\simeq0.378$. For a fitted horizon--seed model, $g_i$ is constant across dates, samples, input positions, and blocks. Thus, adaptation over dates comes from the rolling prior and input-dependent attention, not from changes in $g_i$. Separate horizon fits can learn different allocation vectors.

Two spatio-temporal blocks apply temporal attention, hybrid spatial value aggregation, and a two-layer GELU feed-forward network with a fourfold width expansion. Each operation is followed by a dropout residual connection and layer normalization. The spatial attention has a single 32-dimensional query/key representation. The final calendar-position state, $h_{o,i}^{\rm end}$, supplies the correction head.

\subsection{HAR anchoring and bounded residual fusion}
The horizon-specific HAR coefficients are estimated before neural fitting and remain frozen within that fit. Let $b_{t|o,i}=\max(\widehat y^{\rm HAR}_{t|o,i},\epsilon)$ with $\epsilon=10^{-6}$. The network produces
\begin{align}
 z_{t|o,i}&=\tanh\{f_z(h_{o,i}^{\rm end})\},\\
 r_{t|o,i}&=\sigmoid\{f_r(h_{o,i}^{\rm end},\bar M_{o,i},M_{o,i})\},\\
 \alpha_i&=a_{\max}\tanh(\eta_i),\qquad a_{\max}=1,
 \label{eq:correctionparameters}
\end{align}
where $\bar M_{o,i}$ is the observed fraction of the 22-position input window. Both heads use a 16-unit GELU hidden layer. The last residual-gate bias starts at $-1.5$. The output is
\begin{equation}
 \widehat y_{t|o,i}=
 \max\left[\epsilon,\spplus\left\{
 \spplus^{-1}(b_{t|o,i})+\alpha_i r_{t|o,i}z_{t|o,i}
 \right\}\right].
 \label{eq:fusion}
\end{equation}
The scale $\alpha_i$ is signed. This allows the correction to increase or decrease the benchmark forecast. Initializing $\eta_i=0$ makes the initial forecast equal to the positive HAR anchor, up to floating-point precision. The inverse map is evaluated as $b+\log[-\operatorname{expm1}(-b)]$ for numerical stability.

The correction has a direct mathematical bound. Since $|r z|\leq1$ and softplus is 1-Lipschitz,
\begin{equation}
 |\widehat y_{t|o,i}-b_{t|o,i}|\leq |\alpha_i|\leq a_{\max}.
 \label{eq:bound}
\end{equation}
To obtain Equation~\eqref{eq:bound}, apply the mean-value theorem to softplus, whose derivative is $\sigmoid(x)\in(0,1)$, and use $|\alpha_i r z|\leq|\alpha_i|$. The output floor cannot increase the distance from $b_{t|o,i}\geq\epsilon$. Thus, the bound controls the neural adjustment in the fitted volatility units while leaving the HAR level unrestricted. At initialization, the scale has derivative
\begin{equation}
 \left.\frac{\partial\widehat y_{t|o,i}}{\partial\eta_i}\right|_{\eta_i=0}
 =(1-e^{-b_{t|o,i}})\,a_{\max}r_{t|o,i}z_{t|o,i}.
 \label{eq:initgradient}
\end{equation}
The scale can therefore move away from zero even when the initial forecast equals HAR. Gradients into the residual heads initially carry the factor $\alpha_i=0$ and become active as this scale changes. This parameterization combines an exact benchmark starting point with a trainable, bounded adjustment.

\subsection{Node-balanced MSE with a smooth HAR-relative term}
For a training batch $\mathcal B$, define the observed node loss and active-node set as
\begin{equation}
 \ell_i(\mathcal B)=
 \frac{\sum_{t\in\mathcal B}M_{t,i}(\widehat y_{t|o,i}-y_{t,i})^2}
 {\sum_{t\in\mathcal B}M_{t,i}},
 \qquad \mathcal V_{\mathcal B}=\left\{i:\sum_{t\in\mathcal B}M_{t,i}>0\right\}.
\end{equation}
Let $s_i$ be the standard deviation of observed training targets and $q_i$ the in-sample MSE of the frozen HAR anchor. The implementation floors $s_i$ at $10^{-3}$ and $q_i$ at $10^{-6}$. The objective is
\begin{align}
 \mathcal L(\mathcal B)&=
 \frac{1}{|\mathcal V_{\mathcal B}|}\sum_{i\in\mathcal V_{\mathcal B}}
 \frac{\ell_i(\mathcal B)}{s_i^2}
 +\lambda\,\tau\log\left[
 \frac{1}{|\mathcal V_{\mathcal B}|}\sum_{i\in\mathcal V_{\mathcal B}}
 \exp\left\{\frac{\ell_i(\mathcal B)/q_i-1}{\tau}\right\}\right],
 \label{eq:loss}\\
 &\hspace{2cm}\lambda=0.1,\qquad\tau=0.2.\nonumber
\end{align}
The first term balances markets by their training volatility scale. In the second term, the derivative with respect to $\ell_i$ is $\lambda\pi_i/q_i$, where $\pi_i$ is the softmax weight of $(\ell_i/q_i-1)/\tau$. Markets with larger HAR-relative errors therefore receive greater marginal optimization weight. The term can be negative when relative losses are below their HAR references. All normalizers are estimated within the fitting sample and recomputed in the refit stage. The primary training criterion remains MSE on volatility; QLIKE is an out-of-sample evaluation measure.

\section{Data and experimental protocol}
\subsection{Daily range-based volatility and the union calendar}
Daily OHLC observations come from the historical market data API of Yahoo Finance for the S\&P 500, DAX, CAC 40, FTSE 100, OMX Stockholm All-Share, Nikkei 225, KOSPI, and Hang Seng indices. Taking the latest initial observation and the earliest final observation across the eight series yields a common coverage span from 5 March 2013 to 28 September 2026. The union calendar contains 3,531 dates; 2,870 dates have valid volatility observations in all eight markets.

We measure daily volatility using the Parkinson range estimator \citep{parkinson1980}:
\begin{equation}
 y_{t,n}=100\sqrt{\frac{\{\log(H_{t,n}/L_{t,n})\}^2}{4\log2}}.
 \label{eq:parkinson}
\end{equation}
It is a non-annualized daily volatility proxy in percentage points. The high and low must be finite, positive, and strictly ordered. Positive open and close observations, when present, must lie within the reported range. Invalid ranges become missing observations. The forecasting panel is not forward-filled or joined to a different volatility database. A valid range can be retained for forecasting even if the open or close is unavailable; the economic evaluation additionally requires valid open and close prices.

This sample supports a recent, uniformly sourced international comparison. It measures daily range-based volatility rather than realized variance constructed from intraday returns. Under the assumptions underlying the range estimator, the high--low interval contains information about daily variance. The empirical target here is the observed proxy in Equation~\eqref{eq:parkinson}. All input files already contain its square root on the percentage scale, so the model does not apply a second square-root transformation.
\begin{table}[!htbp]
\centering
\footnotesize
\caption{Sample coverage and descriptive statistics}
\label{tab:data}
\begin{papertabular}{lrrrrr}
\toprule
Market & Observations & Missing (\%) & Mean & SD & Maximum \\
\midrule
SPX & 3413 & 3.34 & 0.640 & 0.473 & 6.142 \\
GDAXI & 3439 & 2.61 & 0.757 & 0.478 & 7.069 \\
FCHI & 3470 & 1.73 & 0.735 & 0.457 & 5.424 \\
FTSE & 3427 & 2.95 & 0.683 & 0.464 & 6.913 \\
OMXSPI & 3394 & 3.88 & 0.662 & 0.418 & 5.225 \\
N225 & 3314 & 6.15 & 0.733 & 0.520 & 7.501 \\
KS11 & 3324 & 5.86 & 0.727 & 0.630 & 10.118 \\
HSI & 3338 & 5.47 & 0.833 & 0.460 & 5.890 \\
\bottomrule
\end{papertabular}
\par\smallskip\begin{minipage}{\textwidth}\footnotesize The common span is 5 March 2013 to 28 September 2026 and contains 3,531 union-calendar dates. Statistics describe non-annualized Parkinson volatility in percentage points. Missingness is measured relative to the union calendar; 2,870 dates have valid observations for all eight markets.\end{minipage}
\end{table}

\subsection{Origin-aligned forecasting protocol and model refitting}
The first 60\% of union-calendar observations form the estimation sample, the next 10\% the validation sample, and the final 30\% the test sample. With integer boundaries, these correspond to 2,118, 353, and 1,060 dates. The estimation sample ends on 23 April 2021 and the validation sample on 31 August 2022. The neural models select the best epoch on the validation set, using at most 150 epochs and patience of 20. They are then initialized again and trained on the first 70\% for exactly that selected number of epochs. The HAR anchor, static graph where applicable, and training normalizers are re-estimated on the refit sample.

Let $b$ be an exclusive sample boundary. Evaluation targets begin at $b+h-1$, placing the first forecast origin at $b-1$. This ensures the same origin-side boundary at every lead. Test targets consequently begin on 1 September 2022, 7 September 2022, and 30 September 2022 for $h=1,5,22$, respectively. Neural and linear coefficients remain fixed during testing; rolling graph updates continue using observations available at each new origin. The target is the volatility of a single future date, not the sum or average of returns over the forecast lead.

Each neural specification is fitted with seeds 1--5. The main loss tables first average its five volatility forecasts and then score that ensemble. Averaging forecasts is different from averaging the five individual-seed losses. We report the latter separately to describe fitting variability. OLS models have one deterministic fit. To ensure rigorous cross-model benchmarking, evaluations across all 13 models are synchronized on identical market--origin--date coordinates with valid ground-truth realizations. The resulting all-days panel contains 8,127, 8,096, and 7,966 market--date observations at the three leads. The corresponding common-days panel contains 857, 854, and 841 dates. It restricts the same forecasts to dates observed in all eight markets; it does not train a second set of models.

All splits, forecast leads, and input restrictions operate on union-calendar date indices. While our primary empirical evaluation focuses on statistical loss criteria, Section~7 extends the analysis to economic utility by detailing the execution assumptions for an intraday portfolio allocation strategy.

\subsection{Benchmarks and controlled variants}
The primary comparison includes HAR, VHAR, HAR-KS, GNN-HAR, DCRNN-HAR, Pure-ST, and PGA-Trans-HAR. Linear models retain OLS estimation on the volatility scale. Neural benchmarks use the equal-node, training-standardized MSE component of Equation~\eqref{eq:loss}, without its HAR-relative term. They share the split, seed, epoch-selection, refit, and evaluation conventions. Table~\ref{tab:settings} reports the common training protocol and model-specific settings. Negative OLS volatility predictions are floored at $10^{-4}$; the main anchored path uses $10^{-6}$ internally. The scoring floor for QLIKE is $10^{-4}$ for every model.

The six controlled variants retain the anchored pipeline and change one designated component. \textit{No prior} sets $g_i=1$; \textit{prior only} sets $g_i=0$; \textit{fixed half} sets $g_i=0.5$; and \textit{scalar gate} fits a single shared gate. \textit{No temporal} replaces each temporal-attention output with zero while retaining the residual normalization path. \textit{MSE only} sets $\lambda=0$ in Equation~\eqref{eq:loss}. These variants are fitted with the same training schedule and seed set. Their trainable parameter counts need not be identical, since removing an operation or fixing a gate changes trainable capacity. The design evaluates the contribution of that specific modelling choice within the common system. Source code and evaluation scripts are publicly available on GitHub at \url{https://github.com/cillinzhao/PGA-Trans-HAR}.

\begin{table}[!htbp]\centering\densetable
\caption{Training and implementation settings}\label{tab:settings}
\begin{papertabular}{p{0.32\textwidth}p{0.64\textwidth}}\toprule
Setting & Value \\\midrule
Common neural protocol & Seeds 1--5; batch 64; maximum 150 epochs; patience 20; 60/10/30 split; refit on first 70\% \\
Main attention path & Window 22; width 32; 2 blocks; 4 temporal heads; spatial width 32; dropout 0.1 \\
Main optimization & AdamW; cosine learning-rate schedule; initial rate $5\times10^{-4}$; minimum rate $10^{-6}$; weight decay $10^{-4}$; gradient clip 1 \\
Main loss and correction & $\lambda=0.1$, $\tau=0.2$, $a_{\max}=1$; initial $\eta_i=0$, $\vartheta_i=-0.5$ \\
Main predictive graph & At most 252 union positions; refresh every 20 origins; VAR(2); ridge 5; GFEVD horizon 22 \\
Main parameter count & 35,666 parameters as recorded by the run configuration \\
GNN-HAR & Hidden width 16; graphical lasso penalty 0.08; learning rate 0.002 \\
DCRNN-HAR & Hidden width 16; 2 diffusion terms; 22-position origin graph; VAR(1); ridge 0; GFEVD horizon 22; learning rate 0.001 \\
Pure-ST & Width 32; 2 blocks; 4 temporal heads; dropout 0.1; learning rate 0.0005 \\
Baseline neural objective & Equal-node training-standardized MSE; requested-lead validation \\
Baseline recurrent training & Endpoint plus intermediate decoder targets; no realized-value feedback during forecast generation \\
Risk/exposure calibration & 5 March 2013--31 August 2022; 22 prior active-day scale for historical tails; pretest median for exposure reference \\
Economic settings & $h=1$ exposure; cap 1.5; $\gamma=3$; annualization 252; daily round trip 0, 2, 5, 10 bps \\
\bottomrule\end{papertabular}
\end{table}

\subsection{Forecast losses and statistical inference}
MSE is the primary forecast loss and MAE summarizes absolute error. The secondary variance-scale QLIKE loss is
\begin{equation}
 L_Q(y,\widehat y)=\frac{y^2}{\widehat y^2}
 -\log\left(\frac{y^2}{\widehat y^2}\right)-1.
 \label{eq:qlike}
\end{equation}

We apply QLIKE to the squared target and squared volatility forecast as a complementary variance-scale score \citep{patton2011}. It assesses relative scale errors and assigns a larger penalty to variance underprediction than to comparable proportional overprediction. Both volatility inputs are floored at \(10^{-4}\) before evaluating the log ratio. MSE and MAE assess errors on the volatility scale. For each loss, aggregate performance is the equal-weighted mean of the eight market-specific losses.

For competitor $j$, we define the loss differential as $d_t=L_{j,t}-L_{{\rm PGA},t}$, so that a positive sample mean indicates lower average loss for PGA-Trans-HAR than for competitor $j$. The Diebold--Mariano test \citep{diebold1995} uses the statistic $\overline d/\sqrt{\widehat\Omega/n}$, with a Bartlett HAC estimate \citep{newey1987} and bandwidth
\begin{equation}
 L_h=\max\left\{h-1,\left\lfloor4(n/100)^{2/9}\right\rfloor\right\}.
\end{equation}
Two-sided normal-reference $p$-values are adjusted by the Holm procedure \citep{holm1979}. The adjustment family consists of the six primary competitors for a given market, horizon, metric, and panel. Ablation comparisons form a separate six-variant family. Results called significant below use the adjusted 5\% threshold.

The Model Confidence Set procedure \citep{hansen2011} uses the seven primary models. We apply the range statistic, 1,000 circular block-bootstrap samples, and block length $\max\{2,h,\operatorname{round}(n^{1/3})\}$ at a 10\% test level. Elimination uses the largest studentized pairwise loss contrast. This yields a 90\% set of retained candidates for each market, horizon, metric, and target-date panel. The primary MCS pool excludes the controlled variants, keeping benchmark inference distinct from internal model comparison. 

\section{Empirical forecasting results}
\subsection{Cross-market accuracy and its horizon profile}
Table~\ref{tab:main} summarizes the out-of-sample predictive performance across all 13 specifications. Among the seven primary models, PGA-Trans-HAR has the smallest average MSE at $h=1$ and $h=22$: 0.15875 and 0.24522, compared with 0.16149 and 0.24802 for HAR. These differences represent reductions of 1.70\% and 1.13\%. Average QLIKE is lowest in the primary pool at all three leads, with reductions from HAR of 3.67\%, 0.43\%, and 2.21\%. The resulting evidence is strongest for short-lead accuracy and for retaining a small long-lead improvement over the local benchmark.

At $h=5$, PGA-Trans-HAR's MSE is 0.21402, close to HAR's 0.21390. Its MAE and QLIKE are the lowest among the primary models. At $h=1$, Pure-ST has the lowest primary-pool MAE, while at $h=22$ that distinction belongs to HAR. These rankings distinguish average absolute forecast deviations from errors that receive greater weight under squared loss. The anchored system improves the latter at the short and long leads. QLIKE adds a relative, variance-scale perspective and penalizes substantial variance underprediction particularly strongly. Its consistent improvement over HAR indicates that the additional information is useful under a risk-sensitive scoring rule as well as the training-related MSE criterion.

\begin{figure}[tbp]
\centering\paperimage{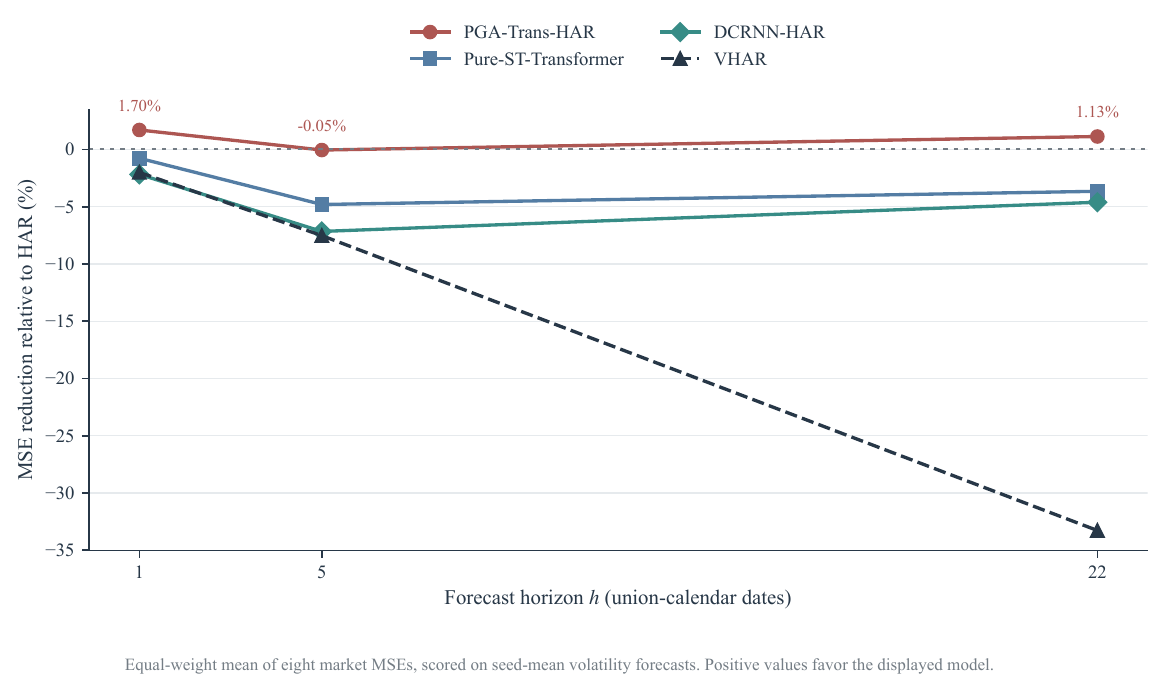}
\caption{MSE gains across forecast leads. Values are $100(1-\overline{\mathrm{MSE}}_j/\overline{\mathrm{MSE}}_{\rm HAR})$ on the all-days panel, with equal weights across market losses. The forecasts are averaged across seeds before scoring. Positive values indicate lower MSE than HAR; the zero line denotes equal loss. Lines connect the three evaluated leads.}
\label{fig:horizons}
\end{figure}

Figure~\ref{fig:horizons} illustrates the term structure of forecast accuracy by plotting the MSE profile across horizons. Pure-ST and DCRNN-HAR have average MSE above HAR at each evaluated lead in this sample. The anchored graph-attention system instead remains close to HAR at the intermediate lead and improves on it at the short and long leads. VHAR's 22-date MSE is 0.33055, compared with HAR's 0.24802. Adding foreign-market predictors therefore does not automatically improve forecasts. The vector specification estimates a broader set of cross-market coefficients, whereas the anchored system organizes those inputs through a predictive graph and limits their correction to the local forecast. The contrast is consistent with a benefit from controlling estimation variability when international signals are noisy.

The horizon profile also clarifies the role of local persistence. HAR's daily, weekly, and monthly components already summarize much of the information relevant at the intermediate lead. The close $h=5$ MSE comparison suggests that the cross-market correction largely preserves this benchmark. At $h=1$ and $h=22$, the correction provides more distinct incremental information. Since each target is a single day's volatility, the long-lead gain concerns the prediction of a distant daily risk state, rather than the accumulation of volatility over a monthly holding period. The economic contribution is thus selective information sharing around a persistent local reference, not a uniformly increasing benefit from a longer forecast horizon.

Table~\ref{tab:marketgains} reports the market-level HAR comparison, while Section~\ref{sec:common} evaluates the same forecasts on synchronized target dates. Together, these comparisons distinguish cross-market heterogeneity from the composition of the target calendar.

\begin{table}[!htbp]
\centering
\densetable
\caption{Equal-market average out-of-sample forecasting losses: all active target days}
\label{tab:main}
\begin{papertabular}{lrrrrrrrrr}
\toprule
 & \multicolumn{3}{c}{$h=1$} & \multicolumn{3}{c}{$h=5$} & \multicolumn{3}{c}{$h=22$} \\
\cmidrule(lr){2-4}\cmidrule(lr){5-7}\cmidrule(lr){8-10}
Model & MSE & MAE & QLIKE & MSE & MAE & QLIKE & MSE & MAE & QLIKE \\
\midrule
HAR & 0.16149 & 0.24258 & 0.36682 & \textbf{0.21390} & 0.27164 & 0.50437 & 0.24802 & \textbf{0.29222} & 0.62402 \\
VHAR & 0.16474 & 0.24402 & 0.38400 & 0.23000 & 0.28534 & 0.54033 & 0.33055 & 0.37926 & 0.67401 \\
HAR-KS & 0.16690 & 0.24519 & 0.37603 & 0.22709 & 0.27737 & 0.53981 & 0.25351 & 0.30224 & 0.62516 \\
GNN-HAR & 0.18512 & 0.26004 & 0.40066 & 0.24700 & 0.29158 & 0.58311 & 0.26251 & 0.29911 & 0.67158 \\
DCRNN-HAR & 0.16502 & 0.24198 & 0.36423 & 0.22922 & 0.27633 & 0.53051 & 0.25946 & 0.29566 & 0.64494 \\
Pure-ST & 0.16274 & \textbf{0.23865} & 0.35632 & 0.22419 & 0.27149 & 0.52637 & 0.25709 & 0.29243 & 0.65270 \\
PGA-Trans-HAR & \textbf{0.15875} & 0.23918 & \textbf{0.35336} & 0.21402 & \textbf{0.27119} & \textbf{0.50219} & \textbf{0.24522} & 0.29267 & \textbf{0.61021} \\
\midrule
no prior & 0.15942 & 0.23939 & 0.35553 & 0.21421 & 0.27117 & 0.50221 & 0.24395 & 0.29137 & 0.60545 \\
prior only & 0.15828 & 0.23956 & 0.35153 & 0.21389 & 0.27153 & 0.50246 & 0.24882 & 0.29260 & 0.62427 \\
fixed half & 0.15925 & 0.23920 & 0.35484 & 0.21443 & 0.27124 & 0.50364 & 0.24368 & 0.29233 & 0.60493 \\
scalar gate & 0.15887 & 0.23924 & 0.35367 & 0.21392 & 0.27098 & 0.50134 & 0.24378 & 0.29208 & 0.60702 \\
no temporal & 0.16048 & 0.24160 & 0.35668 & 0.21461 & 0.27218 & 0.50253 & 0.24808 & 0.29240 & 0.62411 \\
mse only & 0.15967 & 0.23978 & 0.35648 & 0.21465 & 0.27066 & 0.50636 & 0.24729 & 0.29133 & 0.62440 \\
\bottomrule
\end{papertabular}
\par\smallskip\begin{minipage}{\textwidth}\footnotesize Losses are computed from the seed-mean volatility forecast on the common valid model--market--date intersection. Each of the eight markets receives equal weight. MSE is in squared percentage points of volatility; MAE is in percentage points; QLIKE is dimensionless and uses squared targets and predictions. Bold entries identify the smallest loss among the seven primary models above the dividing line, not among the ablations. The six controlled variants below the line retain the HAR-anchored design. Pure-ST denotes Pure-ST-Transformer.\end{minipage}
\end{table}

\subsection{Calendar robustness on common trading days}\label{sec:common}
Table~\ref{tab:common} restricts evaluation to dates with observed targets in all eight markets. Estimation, fitted parameters, and forecasts remain unchanged. Therefore, the comparison measures how the accuracy profile changes with target-calendar composition.

PGA-Trans-HAR retains the lowest primary-pool average MSE at $h=1$ and $h=22$, with reductions of 1.51\% and 1.15\% relative to HAR. It also retains the lowest average QLIKE at every lead, with reductions of 3.69\%, 0.34\%, and 2.31\%. At $h=5$, its MSE is 0.22537 versus 0.22517 for HAR, and its MAE is the lowest among the primary models. The main horizon profile thus carries through to synchronized target dates: short- and long-lead MSE gains, close intermediate-lead MSE, and lower variance-scale loss across all three leads.

This persistence matters for international risk measurement. Restricting a panel to jointly observed dates changes which local risk realizations enter the assessment and removes trading opportunities available in only part of the global market. The similar relative performance on both panels indicates that the forecasting result is not concentrated in these calendar-specific observations. The union-calendar design can retain local trading dates while delivering an accuracy profile that also holds on the conventional synchronized sample.
\begin{table}[!htbp]
\centering
\densetable
\caption{Common-trading-day forecast evaluation}
\label{tab:common}
\begin{papertabular}{lrrrrrrrrr}
\toprule
 & \multicolumn{3}{c}{$h=1$} & \multicolumn{3}{c}{$h=5$} & \multicolumn{3}{c}{$h=22$} \\
\cmidrule(lr){2-4}\cmidrule(lr){5-7}\cmidrule(lr){8-10}
Model & MSE & MAE & QLIKE & MSE & MAE & QLIKE & MSE & MAE & QLIKE \\
\midrule
HAR & 0.16735 & 0.24341 & 0.36692 & 0.22517 & 0.27390 & 0.51422 & 0.25987 & 0.29479 & 0.64271 \\
VHAR & 0.17057 & 0.24514 & 0.38407 & 0.24149 & 0.28748 & 0.54993 & 0.34076 & 0.37976 & 0.68531 \\
HAR-KS & 0.17286 & 0.24590 & 0.37543 & 0.23893 & 0.27973 & 0.55075 & 0.26517 & 0.30461 & 0.64152 \\
GNN-HAR & 0.19148 & 0.26099 & 0.39852 & 0.25994 & 0.29429 & 0.59945 & 0.27487 & 0.30192 & 0.69176 \\
DCRNN-HAR & 0.16948 & 0.24225 & 0.35986 & 0.24156 & 0.27842 & 0.54283 & 0.27259 & 0.29924 & 0.66660 \\
Pure-ST & 0.16877 & 0.23957 & 0.35815 & 0.23560 & 0.27449 & 0.54158 & 0.26911 & 0.29503 & 0.67192 \\
PGA-Trans-HAR & 0.16482 & 0.24012 & 0.35339 & 0.22537 & 0.27357 & 0.51247 & 0.25689 & 0.29514 & 0.62788 \\
\bottomrule
\end{papertabular}
\par\smallskip\begin{minipage}{\textwidth}\footnotesize The panel uses 857, 854 and 841 common target dates at $h=1,5,22$, respectively. Model fitting and forecasts are unchanged. Units and seed aggregation follow Table~\ref{tab:main}.\end{minipage}
\end{table}

\subsection{Paired inference for incremental accuracy}
Table~\ref{tab:dm} summarizes Holm-adjusted paired comparisons. Across the six competitors and eight markets, PGA-Trans-HAR has a significant MSE advantage in 10 of 48 comparisons at $h=1$, 8 at $h=5$, and 9 at $h=22$. None of the all-days MSE comparisons significantly favors the competitor at the adjusted 5\% threshold. The paired tests locate the markets and leads that contribute significant gains.

The HAR comparison is particularly informative. At $h=1$, MSE improvements are significant in SPX, GDAXI, FCHI, FTSE, and OMXSPI. Table~\ref{tab:dmdetail} reports the corresponding statistics and adjusted $p$-values. The concentration of significant short-lead gains in the US and European indices is consistent with the usefulness of shared international information for near-term risk updates. At $h=5$, the significant HAR-relative gains concern FTSE MAE and HSI QLIKE, while KS11 QLIKE favors HAR. At $h=22$, HSI improves significantly under both MSE and QLIKE. The paired evidence therefore supports a market- and horizon-specific use of foreign information: broad short-lead gains coexist with more localized gains further from the forecast origin.

The common-days panel produces 7, 7, and 4 significant MSE advantages across the 48 primary comparisons at the three leads, with no significant MSE disadvantage. The short-lead result therefore persists when all target markets are observed. Changes in significance across panels reflect both the retained dates and the smaller common-day samples. We use these paired tests to locate the gains and MCS to assess the broader set of plausible forecasting candidates.
\begin{table}[!htbp]
\centering
\footnotesize
\caption{Significant paired loss differences relative to PGA-Trans-HAR}
\label{tab:dm}
\begin{papertabular}{lrrrrrrrrr}
\toprule
 & \multicolumn{3}{c}{$h=1$} & \multicolumn{3}{c}{$h=5$} & \multicolumn{3}{c}{$h=22$} \\
\cmidrule(lr){2-4}\cmidrule(lr){5-7}\cmidrule(lr){8-10}
Competitor & MSE & MAE & QLIKE & MSE & MAE & QLIKE & MSE & MAE & QLIKE \\
\midrule
HAR & 5/0 & 6/0 & 5/0 & 0/0 & 1/0 & 1/1 & 1/0 & 0/0 & 1/0 \\
VHAR & 0/0 & 2/0 & 4/0 & 3/0 & 4/0 & 2/0 & 5/0 & 6/0 & 1/0 \\
HAR-KS & 0/0 & 3/0 & 3/0 & 0/0 & 2/0 & 3/0 & 2/0 & 5/0 & 0/0 \\
GNN-HAR & 5/0 & 6/0 & 2/0 & 4/0 & 5/0 & 2/0 & 1/0 & 3/0 & 0/0 \\
DCRNN-HAR & 0/0 & 2/1 & 1/0 & 1/0 & 2/0 & 2/0 & 0/0 & 2/2 & 0/0 \\
Pure-ST & 0/0 & 0/1 & 1/0 & 0/0 & 0/0 & 2/0 & 0/0 & 0/1 & 1/0 \\
\bottomrule
\end{papertabular}
\par\smallskip\begin{minipage}{\textwidth}\footnotesize Each entry reports the number of markets with a significant advantage/disadvantage for PGA-Trans-HAR, out of eight comparisons. Tests are two-sided at 5\%, with HAC standard errors and Holm adjustment across the six primary competitors within each market, horizon and metric on the all-days panel. A non-rejection is not counted in either direction. Table~\ref{tab:dmdetail} details the market-specific MSE statistics against HAR.\end{minipage}
\end{table}

\begin{table}[!htbp]
\centering\footnotesize
\caption{Market-level MSE inference against HAR: all active target days}
\label{tab:dmdetail}
\begin{papertabular}{lrrrrrr}
\toprule
 & \multicolumn{2}{c}{$h=1$} & \multicolumn{2}{c}{$h=5$} & \multicolumn{2}{c}{$h=22$} \\
\cmidrule(lr){2-3}\cmidrule(lr){4-5}\cmidrule(lr){6-7}
Market & DM & $p_{\rm Holm}$ & DM & $p_{\rm Holm}$ & DM & $p_{\rm Holm}$ \\
\midrule
SPX & 3.063 & 0.013 & -0.553 & 1.000 & 1.272 & 0.406 \\
GDAXI & 2.752 & 0.030 & 0.294 & 1.000 & 0.337 & 0.736 \\
FCHI & 3.282 & 0.006 & 0.557 & 1.000 & -0.632 & 1.000 \\
FTSE & 3.505 & 0.002 & 0.639 & 1.000 & -1.692 & 0.272 \\
OMXSPI & 3.544 & 0.002 & 0.747 & 1.000 & 0.122 & 1.000 \\
N225 & 2.476 & 0.080 & 1.579 & 0.572 & 1.715 & 0.345 \\
KS11 & -0.717 & 0.473 & -1.882 & 0.165 & 1.540 & 0.429 \\
HSI & 0.887 & 0.750 & 1.172 & 0.241 & 2.723 & 0.039 \\
\bottomrule
\end{papertabular}
\par\smallskip\begin{minipage}{\textwidth}\footnotesize Positive HAC-DM statistics favor PGA-Trans-HAR. Two-sided $p$-values use the six-competitor Holm family for each market, lead, and metric. The table details the primary MSE criterion; Table~\ref{tab:dm} also summarizes MAE and QLIKE comparisons.\end{minipage}
\end{table}

\FloatBarrier
\subsection{Model Confidence Sets}
Table~\ref{tab:mcs} reports market-level retention counts. On the all-days panel, PGA-Trans-HAR is retained under MSE for 7 of 8 markets at $h=1$ and all 8 at $h=5$ and $h=22$. It is retained under MSE in every common-days market--horizon cell. Across the two panels, this is 47 of 48 MSE cells. Counting all three losses gives 63 of 72 all-days cells and 68 of 72 common-days cells.

Retention is especially consistent at the longer leads. The full model belongs to every 22-date QLIKE set in both panels and every 5-date MAE set. HSI provides a focused example of set discrimination: PGA-Trans-HAR is the sole retained model in its $h=5$ QLIKE set on both panels. It is also the sole candidate in the common-days $h=22$ HSI QLIKE set. Conversely, the all-days SPX $h=1$ MSE set contains DCRNN-HAR and Pure-ST rather than PGA-Trans-HAR. The MCS results thus complement the cross-market losses by identifying where models remain distinguishable or statistically close.

For a forecasting system serving several markets, this distinction is useful. A common architecture benefits from remaining competitive across many local applications, even when another model is preferred for a particular market and loss. PGA-Trans-HAR's broad retention supports that use, while the SPX result leaves room for specialized short-lead alternatives. The larger sets at several longer leads also favor maintaining a pool of forecasting candidates rather than selecting a universal winner from small average-loss differences.
\begin{table}[!htbp]
\centering
\footnotesize
\caption{Retention in the 90\% Model Confidence Set}
\label{tab:mcs}
\begin{papertabular}{lrrrrrrrrr}
\toprule
 & \multicolumn{3}{c}{$h=1$} & \multicolumn{3}{c}{$h=5$} & \multicolumn{3}{c}{$h=22$} \\
\cmidrule(lr){2-4}\cmidrule(lr){5-7}\cmidrule(lr){8-10}
Model & MSE & MAE & QLIKE & MSE & MAE & QLIKE & MSE & MAE & QLIKE \\
\midrule
HAR & 2/3 & 2/3 & 3/2 & 8/8 & 7/7 & 5/5 & 7/7 & 6/7 & 7/7 \\
VHAR & 6/6 & 4/6 & 3/3 & 5/5 & 4/4 & 4/4 & 3/4 & 1/1 & 5/5 \\
HAR-KS & 6/5 & 3/4 & 5/5 & 4/5 & 4/4 & 5/4 & 4/6 & 3/3 & 7/6 \\
GNN-HAR & 2/3 & 2/2 & 4/5 & 3/3 & 3/3 & 5/5 & 6/6 & 4/4 & 7/7 \\
DCRNN-HAR & 7/7 & 5/5 & 6/7 & 7/7 & 5/6 & 5/5 & 8/8 & 6/6 & 8/7 \\
Pure-ST & 8/8 & 8/8 & 7/7 & 8/8 & 8/8 & 5/5 & 8/8 & 6/7 & 7/7 \\
PGA-Trans-HAR & 7/8 & 5/7 & 7/8 & 8/8 & 8/8 & 6/6 & 8/8 & 6/7 & 8/8 \\
\bottomrule
\end{papertabular}
\par\smallskip\begin{minipage}{\textwidth}\footnotesize Entries are counts of retained markets on the all-days/common-days panels, each out of eight. The MCS contains the seven primary models and excludes ablations. Tests use the range statistic and 1,000 circular block-bootstrap draws. Common-day results restrict the same fitted forecasts to target dates observed in all eight markets. Retention identifies candidates whose expected losses remain statistically indistinguishable within the tested model pool.\end{minipage}
\end{table}

\subsection{Market heterogeneity and the role of the local anchor}
Table~\ref{tab:marketgains} shows that the short-lead improvement is distributed across markets. At $h=1$, MSE is lower than HAR in seven of eight markets, while QLIKE is lower in all eight. The largest MSE reduction is 4.53\% for FTSE. At $h=5$ and $h=22$, MSE is lower in six markets at each lead. Long-lead gains reach 2.42\% in HSI and 2.13\% in SPX, while FCHI and FTSE have negative MSE differences. The local comparisons give the cross-market average a concrete interpretation: the gain is broad at short leads and more selective at the long lead.
\begin{table}[!htbp]
\centering\footnotesize
\caption{Market-level loss reductions relative to HAR: all active target days}
\label{tab:marketgains}
\begin{papertabular}{lrrrrrr}
\toprule
 & \multicolumn{2}{c}{$h=1$} & \multicolumn{2}{c}{$h=5$} & \multicolumn{2}{c}{$h=22$} \\
\cmidrule(lr){2-3}\cmidrule(lr){4-5}\cmidrule(lr){6-7}
Market & MSE (\%) & QLIKE (\%) & MSE (\%) & QLIKE (\%) & MSE (\%) & QLIKE (\%) \\
\midrule
SPX & 3.08 & 4.92 & -0.09 & -0.44 & 2.13 & 4.86 \\
GDAXI & 2.54 & 4.87 & 0.10 & 0.79 & 0.23 & 0.66 \\
FCHI & 2.65 & 3.19 & 0.23 & 0.83 & -0.49 & -0.58 \\
FTSE & 4.53 & 6.59 & 0.20 & 0.44 & -1.52 & -1.16 \\
OMXSPI & 2.96 & 3.26 & 0.18 & 0.88 & 0.12 & -0.37 \\
N225 & 3.53 & 4.27 & 0.40 & 1.51 & 1.73 & 2.34 \\
KS11 & -0.54 & 1.13 & -0.74 & -1.99 & 1.26 & 2.87 \\
HSI & 0.64 & 0.39 & 0.56 & 2.50 & 2.42 & 6.54 \\
\bottomrule
\end{papertabular}
\par\smallskip\begin{minipage}{\textwidth}\footnotesize Entries are $100(1-L_{{\rm PGA},i}/L_{{\rm HAR},i})$, calculated from unrounded market-level ensemble losses. Positive values favor PGA-Trans-HAR. Each comparison uses the same valid outcomes within a market and lead. These percentages are market-specific loss ratios; their arithmetic mean differs from the ratio of equal-market average losses in Table~\ref{tab:main}.\end{minipage}
\end{table}

The regional pattern is not uniform across leads. FTSE's large short-lead improvement gives way to a long-lead advantage for HAR, whereas HSI's largest relative gains occur at $h=22$. This contrast suggests that the useful forecasting content of international information has a different time profile across markets. It also explains why a common cross-market architecture should retain a separate local anchor: foreign signals supplement market-specific persistence rather than replacing it with a single global volatility process.

KS11 illustrates the importance of this local reference in a market with relatively dispersed volatility observations. At $h=22$, PGA-Trans-HAR records MSE of 0.69654, compared with 0.70545 for HAR, 0.78522 for Pure-ST, 0.79131 for DCRNN-HAR, and 0.80227 for GNN-HAR. Its QLIKE is 1.01086, compared with 1.04070 for HAR. The difference from the external neural models is larger than the incremental difference from HAR. The economically relevant feature is therefore the preservation of a strong local forecast while allowing a measured adjustment from the international information set.

The profile is lead-dependent. At $h=1$ and $h=5$, HAR has lower KS11 MSE than the full model, while the full model remains substantially below the external neural benchmarks. In the 22-date controlled comparison, fixed-half and scalar-gate variants have KS11 MSE of 0.68527 and 0.68627. Simple allocation rules can therefore exploit the same anchored structure effectively. The market-level evidence places the emphasis on retaining local dynamics and controlling the international correction, rather than on maximizing the flexibility of the gate.

\subsection{Component contributions, fitting stability, and allocation}
The controlled variants in Table~\ref{tab:main} show where the forecast gains arise within the system. At $h=22$, removing temporal attention gives MSE of 0.24808, close to HAR and above the full model's 0.24522. Removing the HAR-relative term gives MSE of 0.24729. At $h=1$, both changes also raise average MSE relative to the full model. Temporal attention allows the correction to distinguish recent observations from older information within the input window. The relative-loss term directs fitting toward improvement over each market's own benchmark. Together, these components address two features of international volatility forecasting: information has different predictive relevance over time, and markets differ in their volatility scale and baseline predictability. The MSE-only variant has the smallest 22-date MAE in the complete 13-model table, showing that the additional objective changes the allocation of errors rather than improving every loss equally.

Graph-only aggregation performs well at shorter leads. The prior-only variant has MSE of 0.15828 at $h=1$ and 0.21389 at $h=5$, slightly below the full model. At $h=22$, its MSE is 0.24882, while the no-prior variant records 0.24395. Fixed-half and scalar-gate fusion achieve 0.24368 and 0.24378. The rolling prior offers a compact summary of recent predictive dependence, which is useful for short-lead signal aggregation. The long-lead comparison favors specifications that give the learned attention more scope alongside the anchor. This pattern is consistent with a horizon-dependent balance between regularized network information and flexible state-dependent inputs. It is the allocation of information, rather than the presence of more connections alone, that matters for accuracy. The forecast-combination findings of \citet{yu2026} similarly emphasize aggregation choices. Here, effective graph-only, fixed-mixture, and learned-mixture specifications share a common HAR anchor; market-specific gating is one implementation of that design.

Table~\ref{tab:seeds} separates ensemble performance from individual-fit variability. At $h=1$, the full model's MSE standard deviation across seeds is 0.00057, compared with 0.01260 for DCRNN-HAR and 0.03727 for GNN-HAR. At $h=22$, the corresponding values are 0.00161, 0.00310, and 0.00764. Lower sensitivity to initialization makes repeated model fitting more reproducible for a risk-forecasting pipeline. The bounded correction supplies an architectural reason for this stability: different neural fits adjust an existing volatility level instead of estimating that level without the frozen reference. The no-temporal variant has even smaller seed dispersion but higher mean MSE, so reproducibility and predictive accuracy remain complementary selection criteria.

\begin{figure}[tbp]
\centering\paperimage{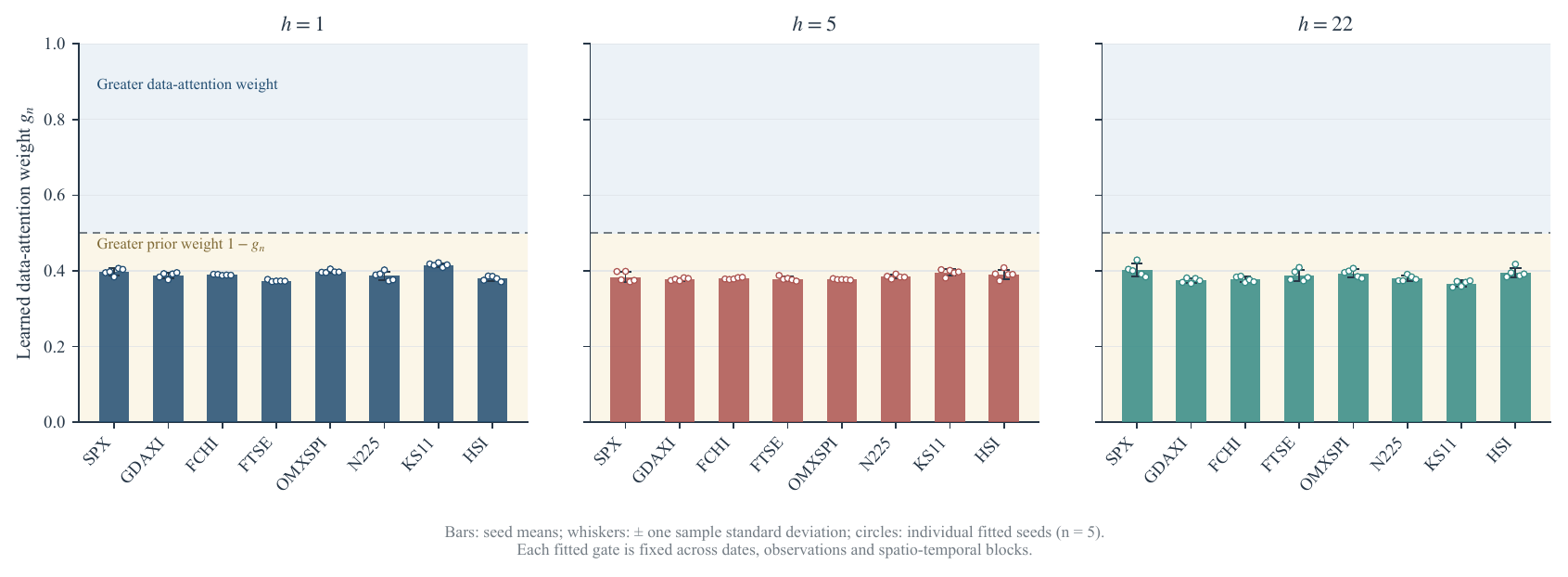}
\caption{Market-specific attention allocation. Bars show five-seed means of $g_i$ at $h=1,5,22$. Whiskers are one sample standard deviation and circles represent individual fits. The gate weights data-driven spatial attention; $1-g_i$ weights the source-masked prior. The 0.5 line indicates equal mixture weights. Each fitted gate is constant over dates and shared across the two spatio-temporal blocks.}
\label{fig:gates}
\end{figure}

Figure~\ref{fig:gates} summarizes the fitted market gates. Their market-averaged seed means are approximately 0.391, 0.383, and 0.384 at the three leads, giving the prior the larger mixture weight. These allocations remain near the initialization of approximately 0.378. They describe a prior-oriented fitted architecture with market-level adjustments, rather than a sharp separation into attention-dominated and prior-dominated markets. The prior retains own-market entries as well as foreign connections, so its weight combines local and international predictive structure. Learned attention adjusts the aggregation to the current input, while the fixed market gate controls its contribution. Date-level changes arise through the rolling prior, neural states, and input-dependent residual gate. This separation permits a changing effective network without requiring a new mixture coefficient for every date.

\begin{table}[!htbp]
\centering
\footnotesize
\caption{Individual-seed MSE: neural benchmarks and controlled variants}
\label{tab:seeds}
\begin{papertabular}{lrrr}
\toprule
Specification & $h=1$ & $h=5$ & $h=22$ \\
\midrule
GNN-HAR & $0.19603\;(0.03727)$ & $0.25808\;(0.03068)$ & $0.26422\;(0.00764)$ \\
DCRNN-HAR & $0.17002\;(0.01260)$ & $0.23196\;(0.00418)$ & $0.26050\;(0.00310)$ \\
Pure-ST & $0.16343\;(0.00083)$ & $0.22457\;(0.00170)$ & $0.25743\;(0.00136)$ \\
PGA-Trans-HAR & $0.15893\;(0.00057)$ & $0.21420\;(0.00048)$ & $0.24562\;(0.00161)$ \\
\midrule
no prior & $0.15968\;(0.00042)$ & $0.21454\;(0.00041)$ & $0.24482\;(0.00244)$ \\
prior only & $0.15858\;(0.00064)$ & $0.21403\;(0.00073)$ & $0.24911\;(0.00198)$ \\
fixed half & $0.15948\;(0.00052)$ & $0.21463\;(0.00071)$ & $0.24431\;(0.00270)$ \\
scalar gate & $0.15905\;(0.00043)$ & $0.21425\;(0.00060)$ & $0.24435\;(0.00268)$ \\
no temporal & $0.16055\;(0.00009)$ & $0.21465\;(0.00044)$ & $0.24811\;(0.00022)$ \\
mse only & $0.16002\;(0.00076)$ & $0.21491\;(0.00061)$ & $0.24774\;(0.00095)$ \\
\bottomrule
\end{papertabular}
\par\smallskip\begin{minipage}{\textwidth}\footnotesize Entries report the mean of the individual-seed equal-market MSEs, with the sample standard deviation across five seeds in parentheses. All seeds are evaluated on the same paired observations. These quantities describe fitting variability and differ from the loss of the averaged forecast in Table~\ref{tab:main}.\end{minipage}
\end{table}

\section{Forecast use: target-day intraday VaR and ES}\label{sec:tailrisk}
The following application evaluates the volatility forecasts as inputs to target-day risk measures. Joint VaR/ES scores assess tail-risk forecasts for unlevered returns under a common calibration, without an exposure cap or transaction-cost rule. Section~\ref{sec:exposureuse} then examines utility under a specified exposure policy. 

\subsection{Construction and evaluation of tail forecasts}
The tail-risk exercise uses open and close prices from Yahoo Finance, the same OHLC source as the volatility panel. The return OHLC files were retrieved on 30 September 2026. Prices and forecasts are matched by market and target date. The outcome is the target-day open-to-close log return,
\begin{equation}
 R_{t,i}^{\rm log}=100\log(C_{t,i}/O_{t,i}),
\end{equation}
and the corresponding ensemble volatility forecast. For each forecast horizon and market, competing specifications are strictly benchmarked on the common intersection of valid forecasts and return realizations. In line with standard practice in daily risk forecasting, the conditional mean return is assumed to be zero. For tail probability $a\in\{0.01,0.05\}$, the signed VaR and ES forecasts are
\begin{equation}
 v_{t|o,i}(a)=\widehat y_{t|o,i}\,q_i(a),
 \qquad e_{t|o,i}(a)=\widehat y_{t|o,i}\,c_i(a),\qquad e<v<0.
 \label{eq:varconstruction}
\end{equation}
Under the benchmark Gaussian tail specification, the parameters are analytically defined as $q_i(a)=\Phi^{-1}(a)$ and $c_i(a)=-\phi(q_i(a))/a$. Historical tails are calibrated separately by market using data from 5 March 2013 to 31 August 2022, before the first test target. The standardized calibration return divides $R_{s,i}^{\rm log}$ by the mean Parkinson volatility over the preceding 22 valid market days, excluding day $s$. The empirical $a$-quantile of these standardized residuals identifies $q_i(a)$, while the truncated sample average below this threshold determines $c_i(a)$. The tail constants are held fixed during testing. This calibration connects the range-based forecast scale to the open-to-close outcome.

We report exceedance frequencies and coverage tests. The Kupiec unconditional-coverage test evaluates the nominal hit probability \citep{kupiec1995}. The Christoffersen conditional-coverage test jointly evaluates coverage and first-order independence of the hit sequence \citep{christoffersen1998}. These tests apply to individual market series. Table~\ref{tab:coverage} reports equal-market mean hit frequencies and the number of market series not rejected by each coverage diagnostic.

Joint VaR/ES scoring uses the FZ0 specification associated with the jointly elicitable pair \citep{fissler2016,patton2019}:
\begin{equation}
 S(R,v,e;a)=-\frac{\indicator\{R<v\}(v-R)}{ae}
 +\frac ve+\log(-e)-1.
 \label{eq:fz}
\end{equation}
Lower scores are preferred for a fixed outcome definition and scoring specification. Each lead concerns the intraday return on a single target date. The volatility forecast supplies the scale, and the fixed tail constants map that scale to VaR and ES.

\subsection{Tail calibration and comparative scores}
Table~\ref{tab:risk} reports the 5\% results for the primary models. At the 1-date lead, PGA-Trans-HAR yields an average historical FZ0 score of 0.61590, compared with 0.62242 for HAR. At the 5-date lead, the corresponding scores are 0.69700 and 0.69919. PGA-Trans-HAR has the lowest historical 5\% score among the primary models at this intermediate lead. At 22 dates, the two historical scores are similar: 0.73745 for PGA-Trans-HAR and 0.73558 for HAR. The historical-tail results therefore concentrate the HAR-relative scoring gains at the short and intermediate leads.

Gaussian-tail FZ0 scores for PGA-Trans-HAR are below HAR at all three leads. At the 1- and 5-date leads, the full model also has the lowest Gaussian 5\% scores in the primary pool. Table~\ref{tab:risk1} reports the 1\% results. At that tail level, historical FZ0 again improves on HAR at the 1- and 5-date leads, while Gaussian FZ0 improves at all three leads. The recurrence of the short- and intermediate-lead pattern at both tail levels gives the forecast improvement a risk-management interpretation. Under a common tail mapping, the adjusted volatility scale produces a better joint assessment of the loss threshold and the severity of losses beyond it.

The historical $h=22$ comparison is also informative. PGA-Trans-HAR improves long-lead QLIKE for the Parkinson target, while HAR has a slightly lower historical FZ0 score at both tail levels. The two evaluations concern different objects: the intraday high--low range and the lower tail of the open-to-close return. A day can have a wide trading range and still close near its opening price. Accuracy for the range-based volatility proxy therefore need not translate one-for-one into accuracy for directional tail losses. The risk exercise shows where the predictive correction transfers to a second financial outcome, with the clearest HAR-relative historical-tail gains at the shorter leads.

Tail calibration also changes exceedance frequencies. The full model's historical 5\% hit rates are 3.380\%, 3.724\%, and 3.266\%, below the nominal level. The Gaussian rates are 6.938\%, 7.215\%, and 6.675\%, above it. Historical 1\% hit rates are 0.530\%, 0.781\%, and 0.835\%. Under the historical mapping, the lower-tail thresholds are more conservative in terms of observed breaches. If these forecasts were used to set risk limits, that conservatism would trade fewer exceedances for tighter exposure limits. The Gaussian mapping makes the opposite trade-off in this sample. Joint scores assess threshold and tail-severity accuracy together, while Table~\ref{tab:coverage} examines the frequency and temporal pattern of breaches.
\begin{table}[!htbp]
\centering
\footnotesize
\caption{Target-day intraday tail-risk forecasts at the 5\% level}
\label{tab:risk}
\begin{papertabular}{lrrrrrr}
\toprule
 & \multicolumn{2}{c}{$h=1$} & \multicolumn{2}{c}{$h=5$} & \multicolumn{2}{c}{$h=22$} \\
\cmidrule(lr){2-3}\cmidrule(lr){4-5}\cmidrule(lr){6-7}
Model & Hit (\%) & FZ0 & Hit (\%) & FZ0 & Hit (\%) & FZ0 \\
\midrule
\multicolumn{7}{l}{\textit{Gaussian tails}} \\
HAR & 6.812 & 0.65127 & 7.129 & 0.76921 & 6.843 & 0.79146 \\
VHAR & 7.233 & 0.65865 & 7.294 & 0.78734 & 4.919 & 0.81131 \\
HAR-KS & 6.959 & 0.65059 & 7.622 & 0.79553 & 6.479 & 0.78416 \\
GNN-HAR & 6.421 & 0.66271 & 6.820 & 0.79768 & 6.773 & 0.82164 \\
DCRNN-HAR & 6.845 & 0.65023 & 6.983 & 0.78566 & 6.736 & 0.80709 \\
Pure-ST & 7.015 & 0.64588 & 7.590 & 0.78939 & 6.965 & 0.81797 \\
PGA-Trans-HAR & 6.938 & 0.63810 & 7.215 & 0.76714 & 6.675 & 0.78472 \\
\midrule
\multicolumn{7}{l}{\textit{Historical tails}} \\
HAR & 3.538 & 0.62242 & 3.711 & 0.69919 & 3.496 & 0.73558 \\
VHAR & 3.748 & 0.61277 & 3.653 & 0.71598 & 2.584 & 0.85011 \\
HAR-KS & 3.604 & 0.61933 & 3.940 & 0.70694 & 3.234 & 0.74965 \\
GNN-HAR & 3.251 & 0.65900 & 3.534 & 0.74680 & 3.385 & 0.75609 \\
DCRNN-HAR & 3.617 & 0.62545 & 3.790 & 0.71420 & 3.405 & 0.75104 \\
Pure-ST & 3.504 & 0.61757 & 3.813 & 0.70187 & 3.506 & 0.74460 \\
PGA-Trans-HAR & 3.380 & 0.61590 & 3.724 & 0.69700 & 3.266 & 0.73745 \\
\bottomrule
\end{papertabular}
\par\smallskip\begin{minipage}{\textwidth}\footnotesize Hit rates and FZ0 scores are unweighted means of the eight market statistics on common valid economic dates. Outcomes are target-day open-to-close log returns in percentage points. Gaussian tails use standard-normal constants. Historical tails use market-specific pretest standardized-return quantiles and tail means. Lower FZ0 is preferable within the same outcome, tail method and horizon. Coverage diagnostics for HAR and PGA-Trans-HAR are summarized in Table~\ref{tab:coverage}.\end{minipage}
\end{table}

\begin{table}[!htbp]
\centering\footnotesize
\caption{Joint VaR/ES scores at the 1\% level}
\label{tab:risk1}
\begin{papertabular}{lrrrrrr}
\toprule
 & \multicolumn{3}{c}{Gaussian tails} & \multicolumn{3}{c}{Historical tails} \\
\cmidrule(lr){2-4}\cmidrule(lr){5-7}
Model & $h=1$ & $h=5$ & $h=22$ & $h=1$ & $h=5$ & $h=22$ \\
\midrule
HAR & 1.26755 & 1.61872 & 1.72298 & 1.03679 & 1.17384 & 1.27401 \\
VHAR & 1.27987 & 1.66032 & 1.61560 & 1.02242 & 1.21200 & 1.39345 \\
HAR-KS & 1.26432 & 1.69473 & 1.68787 & 1.02521 & 1.19668 & 1.28194 \\
GNN-HAR & 1.29951 & 1.71271 & 1.80756 & 1.09796 & 1.27167 & 1.32001 \\
DCRNN-HAR & 1.25664 & 1.67110 & 1.76955 & 1.03197 & 1.20691 & 1.29491 \\
Pure-ST & 1.26102 & 1.68904 & 1.80298 & 1.02616 & 1.19024 & 1.30309 \\
PGA-Trans-HAR & 1.23100 & 1.61482 & 1.69751 & 1.02393 & 1.17225 & 1.27564 \\
\bottomrule
\end{papertabular}
\par\smallskip\begin{minipage}{\textwidth}\footnotesize Entries are equal-market mean FZ0 scores on the same economic-date intersections and under the same calibration rules as Table~\ref{tab:risk}. Lower scores are preferable within each column. Table~\ref{tab:coverage} reports 1\% hit frequencies and coverage diagnostic counts for HAR and PGA-Trans-HAR.\end{minipage}
\end{table}

\FloatBarrier
\subsection{Coverage under alternative tail calibrations}
The historical mapping yields more conservative lower-tail thresholds than the Gaussian mapping in the reported sample. Table~\ref{tab:coverage} illustrates how this changes the hit process. At the 1\% level and $h=5$, both HAR and PGA-Trans-HAR have UC and CC non-rejections in all eight markets under historical calibration. The corresponding Gaussian diagnostics reject in every market. At the 5\% level, the full model's historical UC non-rejection counts are 3, 2, and 2 across the three leads; its CC counts are 2, 3, and 1. Therefore, tail calibration remains consequential even when the volatility forecast is held fixed.

The contrast between the 1\% and 5\% historical diagnostics shows why calibration should be assessed at the intended risk level. A fixed standardized-return distribution can provide satisfactory coverage at one quantile and overly conservative thresholds at another. Unconditional coverage addresses the overall frequency of breaches; conditional coverage also considers whether breaches cluster. Both features matter for a risk limit intended to remain informative over successive trading days. A low average hit rate alone does not establish that the forecast allocates risk accurately over time.

Taken together, Tables~\ref{tab:risk}--\ref{tab:coverage} distinguish two complementary tasks. The volatility model supplies the time-varying scale, while the pretest tail distribution converts it into a threshold and expected shortfall. Both affect joint scores and coverage. Holding the mapping fixed across models isolates their forecast-scale comparison within this exercise. The results favor the anchored correction over HAR in short- and intermediate-lead historical joint scores, and show that tail calibration is a separate operational choice with a material effect on the frequency of risk-limit breaches.
\begin{table}[!htbp]
\centering\footnotesize
\caption{VaR coverage across markets under the two tail calibrations}
\label{tab:coverage}
\begin{papertabular}{lrrrrrrrrr}
\toprule
 & \multicolumn{3}{c}{$h=1$} & \multicolumn{3}{c}{$h=5$} & \multicolumn{3}{c}{$h=22$} \\
\cmidrule(lr){2-4}\cmidrule(lr){5-7}\cmidrule(lr){8-10}
Model & Hit (\%) & UC & CC & Hit (\%) & UC & CC & Hit (\%) & UC & CC \\
\midrule
\multicolumn{10}{l}{\textit{Gaussian tails, 1\% level}} \\
HAR & 2.676 & 1/8 & 1/8 & 3.008 & 0/8 & 0/8 & 2.738 & 1/8 & 0/8 \\
PGA-Trans-HAR & 2.638 & 0/8 & 0/8 & 3.056 & 0/8 & 0/8 & 2.650 & 1/8 & 0/8 \\
\midrule
\multicolumn{10}{l}{\textit{Gaussian tails, 5\% level}} \\
HAR & 6.812 & 2/8 & 4/8 & 7.129 & 2/8 & 2/8 & 6.843 & 3/8 & 3/8 \\
PGA-Trans-HAR & 6.938 & 3/8 & 3/8 & 7.215 & 1/8 & 2/8 & 6.675 & 3/8 & 2/8 \\
\midrule
\multicolumn{10}{l}{\textit{Historical tails, 1\% level}} \\
HAR & 0.517 & 6/8 & 5/8 & 0.781 & 8/8 & 8/8 & 0.849 & 6/8 & 6/8 \\
PGA-Trans-HAR & 0.530 & 5/8 & 7/8 & 0.781 & 8/8 & 8/8 & 0.835 & 6/8 & 6/8 \\
\midrule
\multicolumn{10}{l}{\textit{Historical tails, 5\% level}} \\
HAR & 3.538 & 3/8 & 3/8 & 3.711 & 2/8 & 4/8 & 3.496 & 2/8 & 1/8 \\
PGA-Trans-HAR & 3.380 & 3/8 & 2/8 & 3.724 & 2/8 & 3/8 & 3.266 & 2/8 & 1/8 \\
\bottomrule
\end{papertabular}
\par\smallskip\begin{minipage}{\textwidth}\footnotesize Hit is the equal-market mean exceedance frequency. UC and CC give the number of market series with $p\geq0.05$ in the Kupiec unconditional-coverage and Christoffersen conditional-coverage tests, respectively, out of eight. Tests are applied to each market series, not to pooled or average hits; counts use the unrounded, unadjusted diagnostic $p$-values. Non-rejection is a diagnostic outcome, not proof of correct calibration. All tests are available in these cells. Returns, forecasts, and tail constants match Tables~\ref{tab:risk} and~\ref{tab:risk1}.\end{minipage}
\end{table}

\FloatBarrier
\section{Forecast use: volatility-managed intraday exposure}\label{sec:exposureuse}
This supporting exercise investigates whether the bounded neural correction yields incremental economic utility over the HAR benchmark under a common exposure policy. The comparison holds the return series, pretest reference scale, and cost assumptions fixed across models. HAR-relative CER measures the decision value of the correction, while comparisons with the other models show how exposure levels affect the cost-sensitive rankings.
\subsection{Exposure rules, costs, and utility}
Volatility-managed strategies connect risk forecasts to exposure decisions \citep{moreira2017}. We evaluate eight single-market intraday strategies under a common exposure rule. For each market, let $\sigma_{{\rm ref},i}$ be the median Parkinson volatility from 5 March 2013 to 31 August 2022. This reference scale remains fixed during testing. A forecast from the $h=1$ model determines
\begin{equation}
 w_{t,i}=\min\left\{1.5,\frac{\sigma_{{\rm ref},i}}{\widehat y_{t|t-1,i}}\right\}.
 \label{eq:exposure}
\end{equation}
Gross exposures are established at the target-day market open and liquidated at the close. The constant-exposure comparator also opens and closes daily, with $w=1$. A zero risk-free rate is assumed for uninvested cash during the intraday holding period. Gross and net returns in decimal units are
\begin{equation}
 G_{t,i}=w_{t,i}\left(\frac{C_{t,i}}{O_{t,i}}-1\right),
 \qquad G_{t,i}^{\rm net}(c)=G_{t,i}-10^{-4}c\,|w_{t,i}|,
 \label{eq:costs}
\end{equation}
where $c\in\{0,2,5,10\}$ is the assumed full round-trip cost in basis points. A 5 bps assumption is therefore the total entry-and-exit cost, not 5 bps for each side. Because the strategy carries no overnight positions, transaction frictions are levied on the entirety of the gross daily exposure.

For each market, annualized net Sharpe and mean--variance CER are
\begin{equation}
 \operatorname{SR}_i(c)=\sqrt{252}\frac{\overline G_i^{\rm net}(c)}{s_i^{\rm net}(c)},
 \qquad
 \operatorname{CER}_i(c)=100\times252
 \left[\overline G_i^{\rm net}(c)-\frac\gamma2\{s_i^{\rm net}(c)\}^2\right],
 \quad\gamma=3.
 \label{eq:cer}
\end{equation}
The sample standard deviation uses one degree-of-freedom correction. The reported metrics are equally weighted cross-market averages, summarizing the aggregated performance of the eight single-market strategies. Wealth curves compound simple net returns as $W_t=\prod_{s\leq t}(1+G_s^{\rm net})$.

The exercise compares hypothetical daily index exposures under common cost and cash-return assumptions. Implementation with tradable instruments requires source-market timestamps aligned with each destination-market opening, in addition to the daily date alignment used here.

\begin{table}[!htbp]
\centering
\footnotesize
\caption{Economic evaluation under hypothetical daily round-trip costs}
\label{tab:strategy}
\begin{papertabular}{lrrrr}
\toprule
Model & 0 bps & 2 bps & 5 bps & 10 bps \\
\midrule
\multicolumn{5}{l}{\textit{Panel A: Net Sharpe ratio}} \\
HAR & 0.2503 & -0.1715 & -0.8037 & -1.8560 \\
VHAR & 0.3142 & -0.1082 & -0.7415 & -1.7959 \\
HAR-KS & 0.2969 & -0.1245 & -0.7562 & -1.8080 \\
GNN-HAR & 0.3020 & -0.1088 & -0.7251 & -1.7515 \\
DCRNN-HAR & 0.2761 & -0.1436 & -0.7730 & -1.8209 \\
Pure-ST & 0.2499 & -0.1723 & -0.8053 & -1.8591 \\
PGA-Trans-HAR & 0.2606 & -0.1627 & -0.7974 & -1.8536 \\
Constant exposure & 0.3403 & -0.0174 & -0.5539 & -1.4480 \\
\midrule
\multicolumn{5}{l}{\textit{Panel B: Annual CER (percent)}} \\
HAR & 0.9078 & -3.6104 & -10.3881 & -21.6851 \\
VHAR & 1.5121 & -3.0860 & -9.9835 & -21.4803 \\
HAR-KS & 1.2413 & -3.2784 & -10.0584 & -21.3591 \\
GNN-HAR & 1.0634 & -3.1925 & -9.5764 & -20.2168 \\
DCRNN-HAR & 1.0498 & -3.4212 & -10.1279 & -21.3064 \\
Pure-ST & 0.7739 & -3.7462 & -10.5268 & -21.8284 \\
PGA-Trans-HAR & 0.9867 & -3.5287 & -10.3022 & -21.5922 \\
Constant exposure & 0.2151 & -4.8249 & -12.3849 & -24.9849 \\
\midrule
\multicolumn{5}{l}{\textit{Panel C: Annual CER difference from HAR (percentage points)}} \\
HAR & 0.0000 & 0.0000 & 0.0000 & 0.0000 \\
VHAR & 0.6043 & 0.5244 & 0.4046 & 0.2047 \\
HAR-KS & 0.3335 & 0.3320 & 0.3297 & 0.3259 \\
GNN-HAR & 0.1556 & 0.4179 & 0.8116 & 1.4682 \\
DCRNN-HAR & 0.1419 & 0.1892 & 0.2602 & 0.3787 \\
Pure-ST & -0.1339 & -0.1358 & -0.1387 & -0.1434 \\
PGA-Trans-HAR & 0.0789 & 0.0817 & 0.0858 & 0.0928 \\
Constant exposure & -0.6927 & -1.2145 & -1.9968 & -3.2998 \\
\bottomrule
\end{papertabular}
\par\smallskip\begin{minipage}{\textwidth}\footnotesize The strategy uses $h=1$ forecasts, a maximum exposure of 1.5, the pretest median volatility as its reference scale, and risk aversion $\gamma=3$. Entries are equal-weight averages of market-specific statistics. Annualization uses 252 days. Each strategy, including constant exposure, enters and exits daily. The full round-trip cost applies per unit of daily exposure, and cash earns zero during the holding interval.\end{minipage}
\end{table}

\subsection{HAR-relative utility and exposure-dependent costs}
\begin{figure}[tbp]
\centering\paperimage{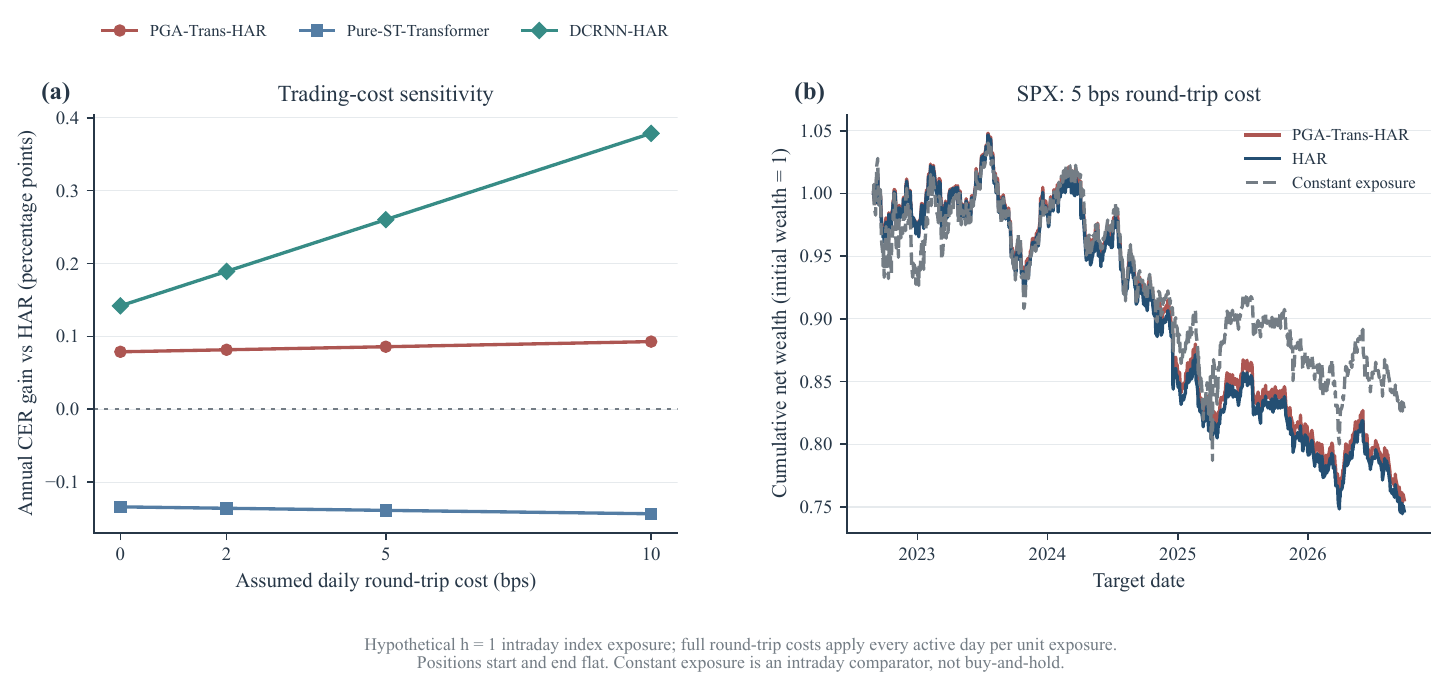}
\caption{CER increments and intraday wealth under a common exposure rule. Panel (a) reports the equal-market mean annual CER difference from HAR at assumed daily round-trip costs of 0, 2, 5, and 10 bps. Differences are annual percentage points. Panel (b) compounds SPX simple net returns at 5 bps for PGA-Trans-HAR, HAR, and constant intraday exposure. Positions are established at the target-day open and liquidated at the close. All series within each market use the same valid dates, and costs apply to the full daily exposure.}
\label{fig:economic}
\end{figure}

Table~\ref{tab:strategy} evaluates whether introducing the neural correction preserves and improves the HAR benchmark's utility under the stated policy. PGA-Trans-HAR has higher average net Sharpe and CER than HAR in all four cost scenarios. Its annual CER increments are 0.07888, 0.08166, 0.08584, and 0.09285 percentage points at 0, 2, 5, and 10 bps. The resulting 7.89--9.28 basis-point annual premium connects the bounded forecast correction to a positive HAR-relative utility increment across the cost grid. Figure~\ref{fig:economic} reports the cost profile and illustrates the corresponding SPX wealth paths at 5 bps.

Without costs, average annual CER is 0.98671\% for PGA-Trans-HAR and 0.90783\% for HAR. Average annualized gross returns are 2.724\% and 2.656\%, with average annualized volatilities of 10.729\% and 10.767\%, respectively. Consequently, these results combine a higher average gross return with lower average volatility under the common exposure rule. Mean exposure is nearly unchanged: 0.89582 for PGA-Trans-HAR and 0.89637 for HAR. The positive utility increment accompanies a reallocation of exposure across dates, rather than greater average market participation. This is the economically relevant comparison for an architecture designed to adjust a strong local forecast.

The cost profile in Figure~\ref{fig:economic}(a) is mechanically driven by the structural friction of daily entry and exit. Annualized mean costs equal $100\times252\times10^{-4}c\,\overline w_i$. PGA-Trans-HAR's slightly lower exposure reduces this cost burden relative to HAR, while costs also affect the variance term in Equation~\eqref{eq:cer}. Together, these effects raise the HAR-relative CER increment as the assumed cost increases. At 5 bps, average CER is $-10.30223\%$ for PGA-Trans-HAR and $-10.38807\%$ for HAR. The levels reflect the cumulative cost of daily round trips, whereas their difference measures the value of changing the forecast within the same policy.

Cross-model CER rankings also reflect differences in the exposure charged for daily round trips. VHAR has the highest primary-pool CER at zero and 2 bps, while GNN-HAR leads at 5 and 10 bps. GNN-HAR's mean exposure is 0.84439, compared with 0.89582 for PGA-Trans-HAR. At 5 bps, this difference lowers GNN-HAR's annualized mean cost by approximately 0.648 percentage points. Its CER advantage over PGA-Trans-HAR increases from 0.077 percentage points without costs to 0.726 percentage points at 5 bps. Much of this widening gap is thus accounted for by the lower mean cost burden. The high-cost ranking reflects both the return--risk allocation and the amount of exposure charged for daily round trips.

The HAR comparison links the proposed architecture directly to its allocation use: under the common rule, the bounded correction yields a positive CER increment across the specified cost scenarios at nearly unchanged average exposure. Together with the joint VaR/ES results, this connects the forecasting improvement to two uses of the estimated risk scale: tail-risk assessment and exposure allocation.

\FloatBarrier
\section{Conclusion and practical implications}
\subsection{Main findings}
This paper studies econometric anchoring as a way to make cross-market neural volatility forecasts more reliable. PGA-Trans-HAR preserves a fitted HAR forecast and learns a bounded correction using masked temporal attention and an origin-aligned predictive graph. A node-balanced objective directs training toward errors relative to each market's scale and HAR reference. The architecture combines local persistence with flexible international information sharing.

The eight-market experiment supports this design through accuracy, fitting stability, and broad MCS retention. PGA-Trans-HAR achieves the lowest average primary-pool MSE at the 1- and 22-date leads and the lowest average QLIKE at all three leads. Its MSE improvements relative to HAR are 1.70\% and 1.13\% at the two endpoints. Paired tests locate significant gains in several markets, and the model is retained in 47 of 48 MSE confidence-set cells across two panels. Crucially, PGA-Trans-HAR exhibits markedly lower initialization sensitivity compared to complex alternatives like GNN-HAR and DCRNN-HAR across all horizons. Fixed and shared allocation rules also deliver accurate forecasts within the anchored architecture. Collectively, this body of evidence substantiates the efficacy of bounded neural corrections, offering a scalable framework to exploit cross-market dynamics under varying degrees of structural flexibility.

\subsection{Implications for forecast users}
For industry practitioners, the proposed design delivers a modular architecture that integrates international spillovers without supplanting robust local dynamics. Source masks retain valid observations across different calendars. The rolling network organizes predictive inputs, and the anchor provides the neural adjustment a stable reference. These features can be incorporated into a forecasting pipeline whose components are separately auditable.

The downstream applications translate statistical forecast improvements into joint VaR/ES evaluations and forecast-based intraday index-exposure rules. Historically calibrated joint risk scores improve on HAR at the 1- and 5-date leads. Under the common exposure rule, annual CER gains relative to HAR range from 7.89 to 9.28 basis points across the stated cost scenarios. These results show how forecast differences carry through to risk scores and exposure decisions. Ultimately, these economic evaluations complement the statistical evidence of accuracy and structural stability, confirming the proposed architecture as both an econometric advancement and a viable tool for global financial practice.

\subsection{Further research}
Future work can examine the framework in larger equity cross-sections, continuously traded markets, and settings with alternative econometric anchors. Incorporating microstructurally synchronized, timestamp-aligned inputs represents a vital engineering extension to facilitate fully actionable, latency-sensitive cross-market execution. Enriching the predictive input space with high-frequency intraday realized measures and sign-asymmetric components would provide a richer testbed for assessing the bounded neural adjustment. Transitioning from univariate volatility paths to full-dimensional joint covariance forecasting stands as a compelling frontier, naturally bridging single-asset exposure rules with global Markowitz portfolio allocation and dynamic cross-border hedging. Importantly, these advances all uphold the core doctrine demonstrated in this study: embedding structured network topology within an econometric anchor effectively unlocks the expressive power of self-attention for high-noise financial markets.
\bibliographystyle{apalike}
\bibliography{references}
\end{document}